\documentclass{emulateapj}
\newcommand{\numobssnr}{15}
\newcommand{\nummesnr}{20}

\newcommand{\kms}{ km s$^{-1}$}
\newcommand{\dsec}{\hbox{$.\!\!^{\rm s}$}}             % Second of time
\newcommand{\damin}{\hbox{$.\!\!^{\prime}$}}           % Arcseconds over dot
\newcommand{\beam}{beam$^{-1}$}
\newcommand\noh{N_\mathrm{OH}}
\newcommand\tex{T_\mathrm{ex}}
\newcommand\ut[2]{\,\textrm{#1}^{#2}}

\newcommand\hi{\textrm{HI}}
\newcommand\hii{\textrm{HII}}

\slugcomment{Accepted to ApJ, 26 March 2008}
\shorttitle{Survey of Hydroxyl in Supernova Remnants}
\shortauthors{Hewitt, Yusef-Zadeh \& Wardle}

\begin{document}
\title{A Survey of Hydroxyl Toward Supernova Remnants: Evidence for Extended 1720 MHz Maser Emission}
\author{J.W. Hewitt, F. Yusef-Zadeh}
\affil{Department of Physics and Astronomy, Northwestern University, Evanston, IL 60608; j-hewitt@northwestern.edu}
\and
\author{M. Wardle}
\affil{Department of Physics, Macquarie University, Sydney, NSW 2109, Australia}

\begin{abstract}
We present the results of GBT observations of all four ground-state hydroxyl(OH) transitions toward \numobssnr\ supernova remnants(SNRs) which show OH(1720 MHz) maser emission. This species of maser is well established as an excellent tracer of an ongoing interaction between the SNR and dense molecular material. For the majority of these objects we detect significantly higher flux densities with a single dish than has been reported with interferometric observations. We infer that spatially extended, low level maser emission is a common phenomenon that traces the large-scale interaction in maser-emitting SNRs. Additionally we use a collisional pumping model to fit the physical conditions under which OH is excited behind the SNR shock front. We find the observed OH gas associated with the SNR interaction having columns $\noh \le$ 1.5$\times$10$^{17}$ cm$^{-2}$, temperatures of 20--125 K, and densities $\sim$10$^5$ cm$^{-3}$. 
\end{abstract}

\section{Introduction}

Understanding the interaction between supernova remnants(SNRs) and molecular clouds is crucial in determining the structure and physical conditions that constitute the Galactic interstellar medium. Shocks driven into adjacent clouds inject energy, change the chemical evolution, and possibly trigger star formation within the cloud. Very few clear cases of such interactions had been identified until \citet{frail94} studied masing of the 1720 MHz satellite line of hydroxyl(OH) associated with shocked molecular clumps at the interface between the supernova remnant W28 and an adjacent molecular cloud \citep{wooten81}. In this and subsequent studies it became clear that narrow OH(1720 MHz) masers with corresponding broad OH absorption at 1667, 1665 and 1612 MHz trace the edges of molecular clumps shocked by the passage of the expanding SNR. 

\begin{figure} \centering
\includegraphics[angle=0,width=6in]{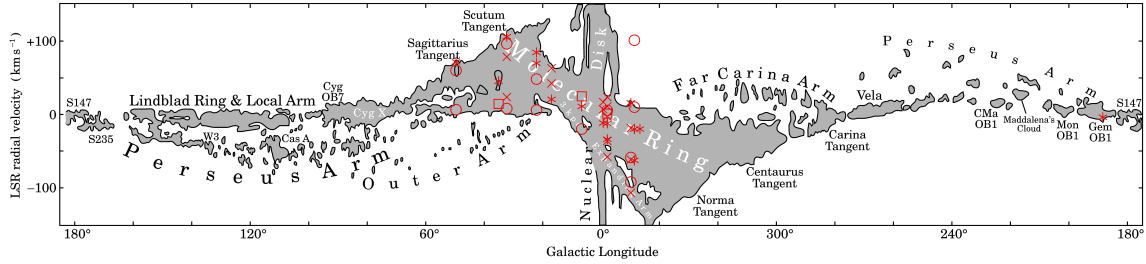}
\caption{Galactic longitude vs. velocity diagrams showing the distribution of OH spectral components we have identified toward each of the observed SNRs superposed on figure 3 from Dame et al. (2001) which shows the Galactic molecular gas distribution traced by CO.
Asterisks indicate OH(1720 MHz) maser emission associated with interacting supernova remnants; open circles indicate anomalous 1720 MHz clouds; open squares indicate anomalous 1612 MHz clouds; and crosses indicate main-line OH absorption features which are not associated with any of the other previously listed physical phenomenon.
\label{fig:galdia}}
\end{figure}

The presence of OH(1720 MHz) masers without inversion of the other ground-state transitions at 1667, 1665 and 1612 MHz is now established as a reliable signpost of interacting SNRs. Efforts to survey known SNRs has resulted in positive detections toward \nummesnr\ Galactic SNRs as well as extragalactic SNR N49 in the LMC, consistent with $\sim$10$\%$ of remnants harboring associated masers \citep{green97,fyz99,brogan04,roberts05}. As can be seen in Figure $\ref{fig:galdia}$, maser-emitting SNRs are preferentially found within the Molecular Ring, with some evidence of an enhanced occurance within the Nuclear Disk \citep{green97}.

Each interacting remnant is an excellent laboratory for understanding shock physics and a valuable target for further study with complementary radio, X-ray, sub-mm, IR and optical observations. Theoretical considerations strongly constrain conditions under which only the 1720 MHz line can be inverted. Collisional pumping in the absence of strong FIR can only produce OH(1720 MHz) masers in molecular gas with temperatures of 15--200 K and densities of $\sim$10$^5$ cm$^{-3}$ \cite{lockett99}. Thus in the absence of the 1665/7 MHz transitions, the OH(1720) MHz line traces cooling post-shock gas.

As the shock propagates into the adjacent cloud, shock chemistry predicts that OH is rapidly converted to water within the shock. 
\citet{wardle99} found that a secondary FUV flux arises when the X-rays emitted by the SNR are incident upon the adjacent dense molecular material.
When the shocked molecular gas has cooled below 180 K, this secondary FUV flux can efficiently dissociate shock-produced water into OH, naturally producing OH at the temperature and abundance required for collisional pumping of the 1720 MHz transition.
The OH enhancement (OH/H$_2$$\ga$10$^{-6}$) which arises behind the shock can be detected as thermal absorption against the bright background continuum of the remnant.

High resolution study of compact maser spots has provided an important means of measuring the post-shock magnetic field. Zeeman splitting of the maser line permits the line of sight field strength to be derived, yielding field strengths of order 1 mG for nearby remnants \citep{claussen97,brogan00} but as high as 5 mG for Sgr A East \citep{fyz96}. VLBI studies have resolved the size and geometry of bright masing clumps towards W28, W44 and IC443 \citep{claussen97,hoffmanphd}.

Additionally, it has become increasingly apparent that weak, extended OH(1720 MHz) maser emission is also present on parsec scales in SNRs W28, G359.1-0.5, G357.7+0.3 and G357.7-0.1 (Yusef-Zadeh et al. 1999, 2003). This extended emission is well correlated with shocked molecular tracers (Lazendic et al 2002) and OH absorption in the other ground-state transitions. Motivated by this unique opportunity to probe very specific conditions just behind the shock front we have observed the four ground-state transitions of OH for all maser-emitting remnants which can be observed in the N. Hemisphere. 

The results for each remnant are discussed in turn in Section $\ref{sec:results}$. The evidence and properties of extended maser emission are discussed Section $\ref{sec:eme}$ with modeling of the physical conditions for the post-shock OH gas is presented in Section $\ref{sec:modeling}$. We differentiate between SNR-type masers and the OH satelite line inversions seen towards large clouds cataloged by \citet{turner82}. Discussion of these "anomalous OH clouds" is given in Section $\ref{sec:anom}$.

\section{Observations and Data Analysis}
\subsection{Observations}
Motivated by the discovery of large-scale OH(1720 MHz) emission toward prominent interacting supernova remnants and the chance to observe all four transitions simultaneously we used the the Green Bank Telescope (GBT) of the National Radio Astronomy Observatory\footnote{The National Radio Astronomy Observatories is a facility of the National Science Foundation, operated under a cooperative agreement by Associated Universities, Inc.} to survey \numobssnr\
maser-emitting supernova remnants. Observations were conducted from August to October of 2003 as part of proposal GBT02B-006. All four ground-state transitions of OH at 1612.231, 1665.4018, 1667.359 and 1720.530 MHz in left- and right-circular polarizations were simultaneously observed.  

Table \ref{obstbl} lists the observed SNRs and the extent of our mapping. For all but G359.1--0.5, the full extent of each remnant was observed via point mapping where each pointing is spaced by 3$\damin$3 corresponding to the Nyquist sampling limit for the GBT beam-size of 7$\damin$2 at 1.7 GHz.
Each pointing is observed for a time sufficient to detect all previously reported OH(1720 MHz masers).

Spectra taken with the auto-correlation spectrometer have a velocity resolution of 0.28\kms\ over a 12.5 MHz bandwidth. In-band frequency switching of 1.17 MHz is employed for calibration limiting effective velocity coverage to 400\kms\ about the rest frequency of each line.
Data reduction was performed with GBTIDL. A low-order polynomial fit was subtracted from the full baseline to remove ripples in the spectrometer baselines. For modeling it was necessary to obtain the continuum level from our observations. The difference of the system temperature measured on source compared to our blank sky position ($\Delta$T$_{sys}$) is used to derive the continuum temperature (T$_{src}$) using the equation:
\begin{equation}
\Delta T_{sys} = 3 \Delta A + \eta  T_{src}  {\rm exp}(\tau A_{src})
\end{equation}
where $\eta$ is beam efficiency (0.94), $\tau$ is atmospheric opacity (0.011 at L-band), and $\Delta$A = 1/sin($\Theta_{src}$)--1/sin($\Theta_{sky}$) where $\Theta$ is the elevation of observations on source and on blank sky. We estimate that the obtained continuum temperature is accurate to 10$\%$.

\subsection{OH Excitation Modeling \label{sec:modelfit}}
Observations of all four ground-state transitions permits detailed modeling of the post-shock OH gas in which the masers arise. Using the collisional pumping model of \citet{wardle07} we are able to estimate the parameters of the gas responsible for producing the observed line profiles using a simple multi-component model. This modeling procedure has already been successfully demonstrated for SNR IC443 (see Hewitt et al. 2006 for further details). 

The observed OH spectra are modeled using both narrow components for maser emission or a broad components to fit absorption from shocked gas. Each component is characterized by its line-of-sight OH column density $\noh$, kinetic temperature $T_k$, mean and FWHM velocities, $v_0$ and $\Delta v$, and a beam filling factor $f$. The H$_2$ number density is in almost all cases found to be $10^5$\,cm$^{-3}$, typical of what is expected in OH(1720 MHz) maser-emitting regions.
Excitation by collisions and partial trapping of photons then determines the excitation temperature $\tex$ of the four transitions and the line-center optical depth. 

The observed intensity for a particular line is
\begin{equation}
     I_\nu=(1-f)I_{0\nu}+f\left[I_{0\nu}\exp(-\tau_\nu)+
     B_\nu(\tex)(1-\exp(-\tau_\nu))\right]
     \label{eq:Inu}
\end{equation}
where $I_{0\nu}$ is the background continuum intensity and $B_\nu$ is
the Planck function. Here the possible obscuration of one component by another 
is neglected. Assuming that $|\tex|\gg 0.08$\,K, the
Rayleigh-Jeans approximation holds, and eq. (\ref{eq:Inu}) can be recast as
\begin{equation}
     T(v) - T_0 = f (\tex - T_0 )  (1 - \exp(-\tau(v)))
     \label{eq:T_v}
\end{equation}
We write the optical depth as
\begin{equation}
     \tau(v) = \tau_0 \exp\left(-\frac{(v-v_0)^2}{2\sigma^2}\right)
     \label{eq:tau_v}
\end{equation}
where $\tau_0$ is the optical depth at line center and $\sigma = \Delta
v / (8 \ln 2)^{1/2}$ and we have adopted a Gaussian line profile.

The optical depth at line center can be written as 
\begin{equation}
	\tau_0 = a\,
	\frac{\noh(10^{15}\ut{cm}{-2})}{T_{\mathrm{ex}}(\textrm{K}) \Delta
	v(\textrm{km/s})}
	\label{eq:tauOH}
\end{equation}
where the constant $a$ is 0.4540, 2.3452, 4.2261, and 0.4846 for the 1612, 1665, 1667 and 1720\,MHz lines respectively.
%as the ground rotational state are very nearly populated according to their degeneracies.
Note that $\tex$ and $\tau_0$ are negative for an inverted transition.

Eqs (\ref{eq:T_v}), (\ref{eq:tau_v}) and (\ref{eq:tauOH}) form the
basis of a simple model for the observed lines, with parameters $\noh$,
$T_k$, $f$, $v_0$, $\Delta v$.  The excitation temperatures are
determined by an LVG model of the collisional excitation of the OH
molecules within a clump, including the effect of line overlap and
radiative transfer in the maser line, following
Lockett, Gauthier \& Elitzur (1999).  

We use the minimum number of components to achieve an acceptable fit to the data. To obtain the parameters $f$, N$_{OH}$, V$_{LSR}$, and $\Delta$V we have utilized the Markwardt IDL fitting routines to minimize our $\chi ^2$ fit. The resulting model fitting including both maser and absorption components is presented in Table $\ref{tbl:modelfits}$. 

\section{Results}\label{sec:results}

In the following section we present results of our OH mapping for each remnant in turn, including discussion of the observed maser emission and broad absorption profiles.
Table $\ref{tbl:eme}$ summarizes the integrated maser-line intensity and velocity range over which maser emission is seen for our sample of remnants. A comparison with VLA detections in the literature allows us to identify which remnants show evidence for maser emission being resolved out by the interferometer. 
Unlike radiatively pumped masers, OH(1720 MHz) masers associated with SNRs are observed to be larger in spatial extent and are not time variable \citep{hoffmanphd}. 
Therefore we infer that the excess flux detected by the GBT is from extended structures and low-gain masers not resolved by previous synthesis imaging. Evidence of extended maser structures has been previously detected with sensitive VLA imaging of bright, large diameter SNRs such as W28 \citep{fyz03_w28}. 

Of the \numobssnr\ observed remnants we find 10 show a significantly higher flux than reported for previous interferometric observations with the VLA. 
For the three SNRs that lie close to the Galactic center foreground absorbing gas makes accurate flux determinations particularly difficult, and so we do not include them in our analysis.
Three of the observed sources (W51C, G348.5-0.0 and CTB 37A) show no discernable change in maser-line emission, consistent with a compact nature for these maser sources. However, for SNRs IC443 and Kes 69 we detect maser emission from regions which were not detected by the VLA, possibly indicating only extended maser emission is present.  Section $\ref{sec:eme}$ gives further discussion of the properties of this inferred extended emission component.
For all areas of interest we present the observed spectra, both over a large velocity range and the velocity range where we see the SNR-cloud interaction. The bold line shows the model and the dashed line shows the residuals from the model fit.

\subsection{G6.4-0.1 (W28)}
W28 is both the first identified maser-emitting SNR and the prototype for its class. At a distance of 2.5 kpc, the remnant has a projected diameter of 30 pc (48\arcmin ). A bright radio shell and centrally concentrated thermal X-ray emission mark W28 as a member of the mixed-morphology class of remnants. A strong correlation is seen between maser-emitting and mixed-morphology remnants, thought to be because both arise through the interaction with an adjacent cloud  \citep{fyz03mmsnrs}. The long ionization timescale and low density of the thermally emitting X-ray plasma indicates an age of order 10$^4$ years for W28  \citep{rho02}. The remnant is coincident with an unidentified EGRET source and has been suggested as a possible site for the origin of $\gamma$-rays \citep{esposito96}.

Where the shock wave from the remnant is impacting the adjacent molecular cloud is delineated by remarkably broad millimeter lines of CO, CS and HCO$^+$ \citep{wooten81,reach05}. Compact OH(1720 MHz) masers show an excellent correlation with shocked molecular filaments as traced by high-resolution CO data \citep{frail98,arikawa99}. Shock speeds of 20-30\kms\ and gas densities $\ga$10$^3$ cm$^{-3}$ are observed consistent with the nondissociative C-type shocks behind which OH(1720 MHz) masers form. \citet{fyz03_w28} observed both broad line width OH(1665/1667 MHz) absorption and extended OH(1720 MHz) emission structures well correlated with the shock. \citet{hoffman05_w28} has resolved the brightest compact maser spots to have linear sizes of only 50 AU. Zeeman measurements of these compact spots indicates magnetic field strengths of $\sim$0.5 mG.

\begin{figure} \centering
\includegraphics[angle=0,width=5.5in]{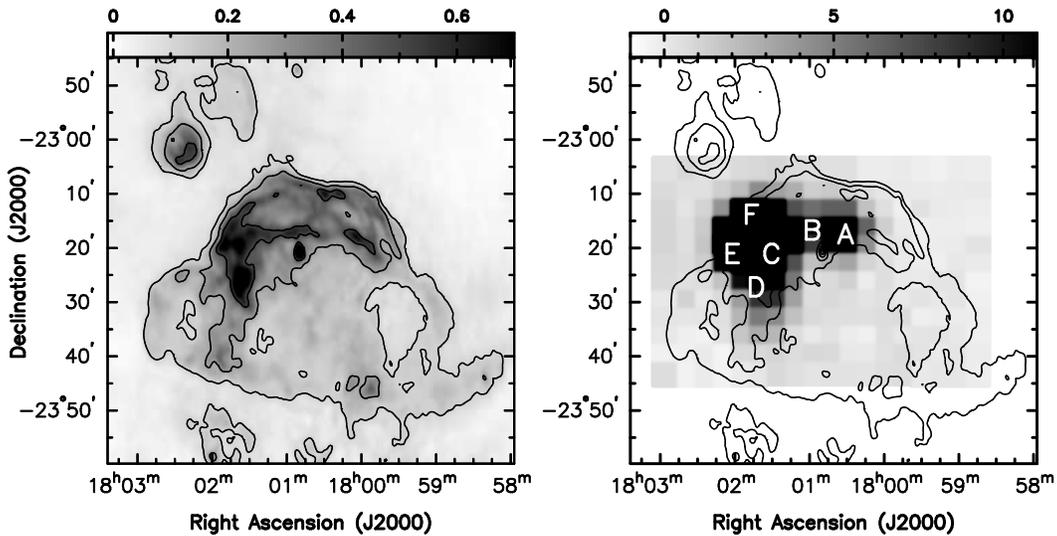}
\caption{[Left] 20cm Radio continuum image of SNR W28 with contours at 0.11, 0.25, 0.5, 1.0 and 1.5 Jy \beam (synthesized beam of 15\arcsec ). [Right] Integrated line intensity of 1720 MHz maser emission towards W28 in units of Jy \kms . Maser groups A through F are labeled. 
\label{fig:mom-w28}}
\end{figure}

Using single dish observations of W28 we detect twice as much flux from 1720 MHz masers as previously reported with VLA A-array observations. The integrated maser-line intensity is shown in Figure $\ref{fig:mom-w28}$ with bright masing clumps A through F identified. The enhanced maser emission is consistent with the detection of extended maser structures corresponding to shocked CO filaments tracing the interaction \citep{fyz03_w28}. These extended maser structures account for as much as half the detected maser-emission. We have modeled the A and E+F maser regions which are spatially distinct in our maps. Best-fit parameters given in Table \ref{tbl:modelfits} are consistent with high columns of $>$10$^{16}$ cm$^{-2}$ and a mix of both small and modest filling factors indicating a mix of compact and extended maser-emitting regions are present.

\begin{figure}
\centering
\includegraphics[angle=-90,width=3.5in]{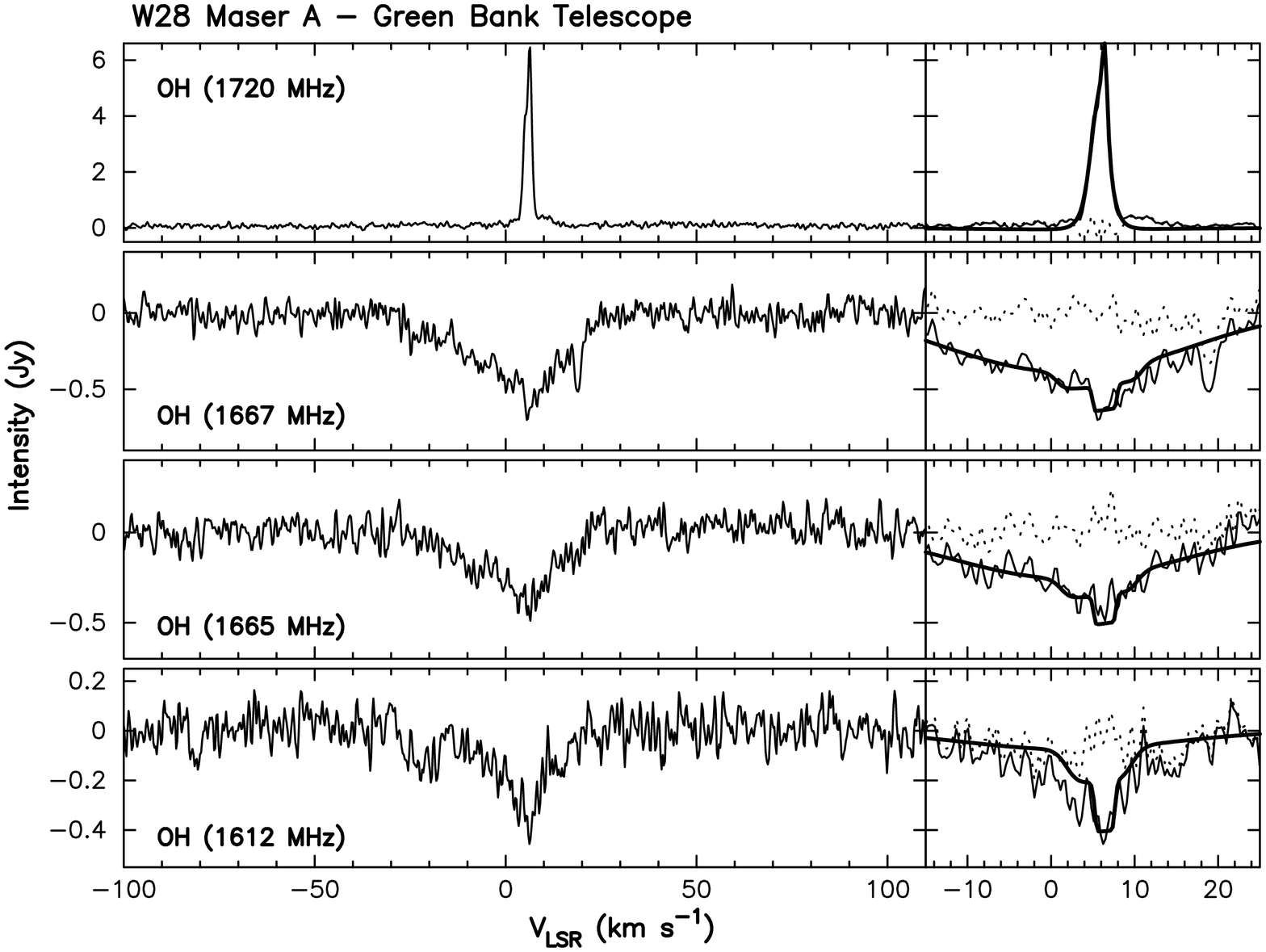}
\includegraphics[angle=-90,width=3.5in]{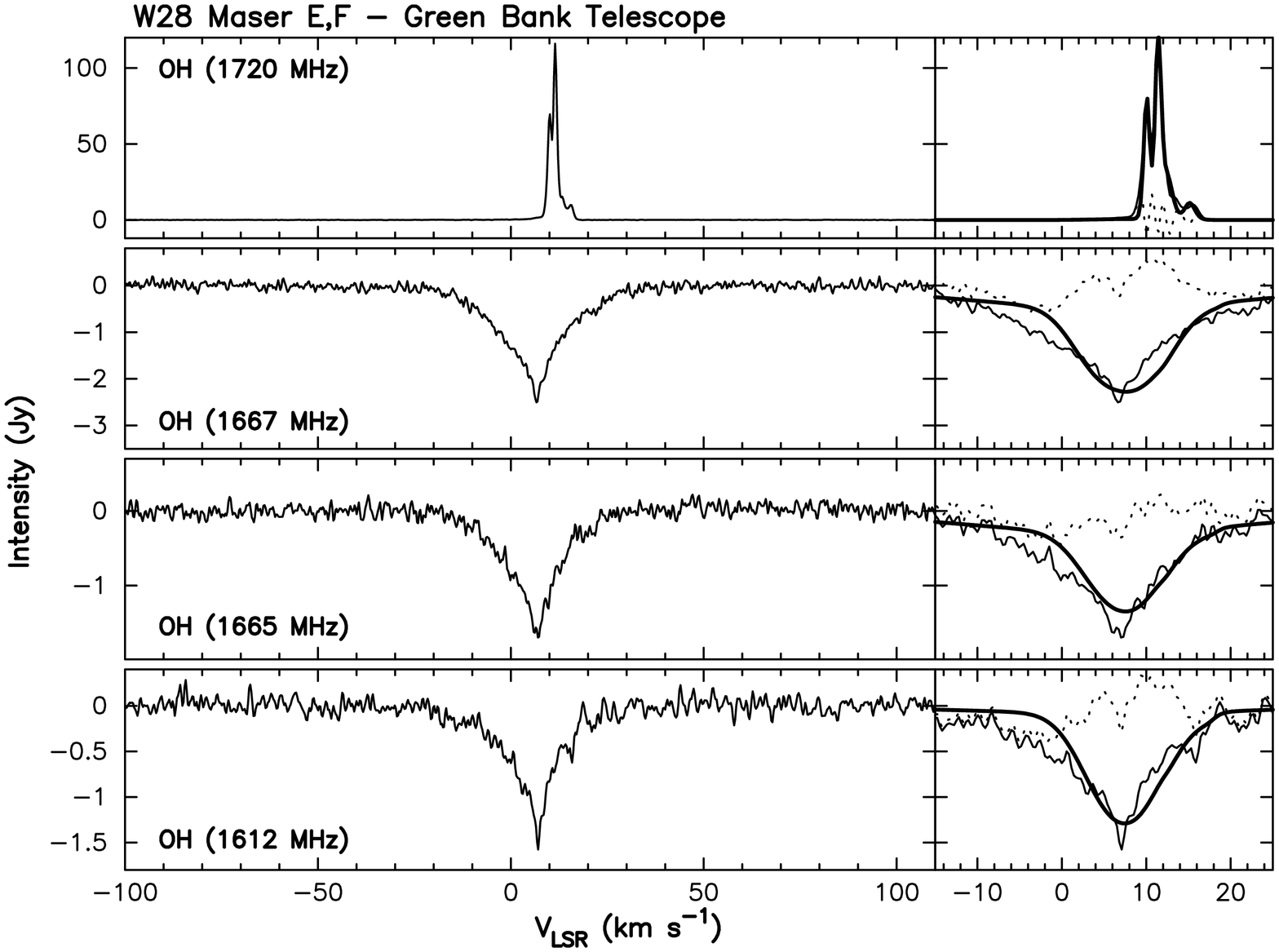}
\caption{Spectra of all four OH transitions towards W28. Each figure shows the observed spectra over a large velocity range and the velocity range where we see the SNR-cloud interaction. The bold line shows the model and the dashed line shows the residuals from the model fit. This is done for all subsequent Figures which present spectra. Maser clump A at the position 18h 00m 44.4s , -23d 16m 53.8s (J2000). Broad symmetric absorption is seen around the maser velocity.
Maser Clumps E and F at the position 18h 01m 50.4s , -23d 19m 13.3s (J2000). Broad symmetric absorption is seen around the maser velocity, but appears offset by a few \kms .
\label{fig:spec-w28}} \end{figure}

Spectra towards the masing regions in Figure $\ref{fig:spec-w28}$ show strong narrow maser emission with broad symmetric absorption with line widths of up to $\sim$40\kms , a clear indication of OH arising in the shocked gas. Symmetric absorption profiles indicate that the shock is propagating transversely across the plane of the sky, whereas strongly blue- or red-shifted profiles indicate the shock is propagating along the line of sight. Bright compact masers are often seen only when the shock is propagating perpendicular to the line of sight as this shock geometry maximizes the pathlength of OH which has a coherent velocity \citep{claussen97}.

We also note instances of conjugate 1720/1612 MHz lines which are kinematically distinct from the W28-cloud interaction. Toward the interior of the remnant we see 1720 MHz emission with conjugate 1612 MHz absorption at --21.5\kms . Along the eastern edge of W28 where the radio continuum is brightest we see 1612 MHz emission at +24.1\kms\ with conjugate 1720 MHz absorption which extends to the southern edge of our map. Linewidths of a few \kms\ are seen, uncharacteristically broad for masers. These are likely molecular clouds which lie along the sight to W28 
as has been seen throughout the galaxy in the survey by \citet{turner82}. This so called "anomalous" emission at 1720 MHz arises in physical conditions distinct from SNR-masers and is discussed further in \S \ref{sec:anom} .

Finally, we note the identification of OH/IR stars which show double-peaked 1612 MHz maser emission. We detect maser emission from the OH/IR star centered at v$_{LSR}$ = +103 \kms\ which has been previously identified \citep{sevenster01}. Also we find an OH/IR star with peaks at --38.3 and --54.2\kms\ which imply an outflow velocity of 9\kms\ and a v$_{LSR}$ of --45.3\kms .

\subsection{G16.7+0.1}

Relatively poorly studied in comparison to other remnants in this work, G16.7+0.1 is a maser-emitting remnant with a composite radio morphology in which the radio shell is superposed with a central pulsar wind nebulae \citep{bock05}. The radio shell appears to brighten along the southern edge where \citet{green97} detected a single maser at +20\kms . 
CO clouds at a velocity of +25\kms\ are present at the location of the maser as well as in the interior and northwestern edge confirming the interaction \citep{reynoso00}. 

\begin{figure} \centering
\includegraphics[angle=-90,width=3.5in]{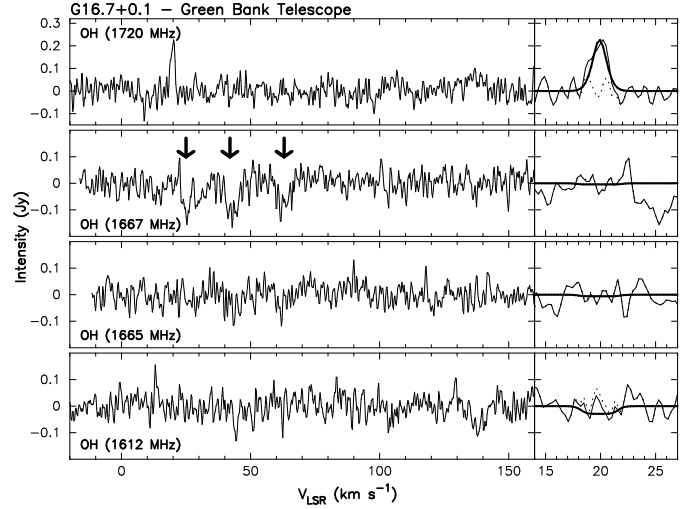}
\caption{Spectra of all four OH transitions towards G16.7+0.1. The maser is seen near +20\kms . Bold arrows indicate OH(1667 MHz) absorption features at +25, +42 and +63 \kms .
\label{fig:spec-g167}}\end{figure}

For the lone detected maser, VLA observations detected a flux of 115 mJy \citep{green97}. GBT observations detect a significantly higher flux of 232$\pm$34 mJy indicating the presence of either additional low-brightness masers not detected by the VLA or extended emission along the southern extent of the SNR. As can be seen in the spectra of Figure \ref{fig:spec-g167}, OH(1667 MHz) absorption also is present at +25\kms\ matching the detected CO. Foreground absorption across the center of the remnant is also seen at +63 and +42 \kms\ for the main-lines, but we do not detect OH absorption at the maser velocity of +20\kms\ ($\sigma_{rms}\sim$50 mJy beam$^{-1}$). As we only have upper limits for the OH absorption associated with the maser, modeling gives an upper limit to the filling factor of 0.005 and column density of 10$^{17}$ cm$^{-3}$ associated with the +20\kms\ maser.

We also note the re-detection of the OH/IR star at v$_{LSR}$ of +26.1 \kms\ \citep{sevenster01}.

\subsection{G21.8-0.6 (Kes 69)}
Kes 69 has an incomplete radio shell morphology with interior and perhaps shell-like X-ray emission detected by Einstein and ROSAT \citep{fyz03mmsnrs}. While the southern ridge of the SNR is bright at 20 and 90 cm \citep{kassim92}, the faint northwestern shell is confused with prominent HII region G21.902--0.368. 
For this HII region recombination line emission gives a velocity of +80\kms \citep{Lockman89}. Kinematics for the SNR are less clear. A single OH(1720 MHz) maser is detected at a velocity of +69\kms\ but its position is roughly 10\arcmin\ to the north of the bright radio shell \citep{green97}. In the northern extent of Kes 69 the shell is not well defined but \citep{green97} argued that the maser is associated with the supernova remnant and not the HII region based on the lack of any main-line OH(1667/5 MHz) maser emission characteristic of HII regions. 

\begin{figure} \centering
\includegraphics[angle=-90,width=3.5in]{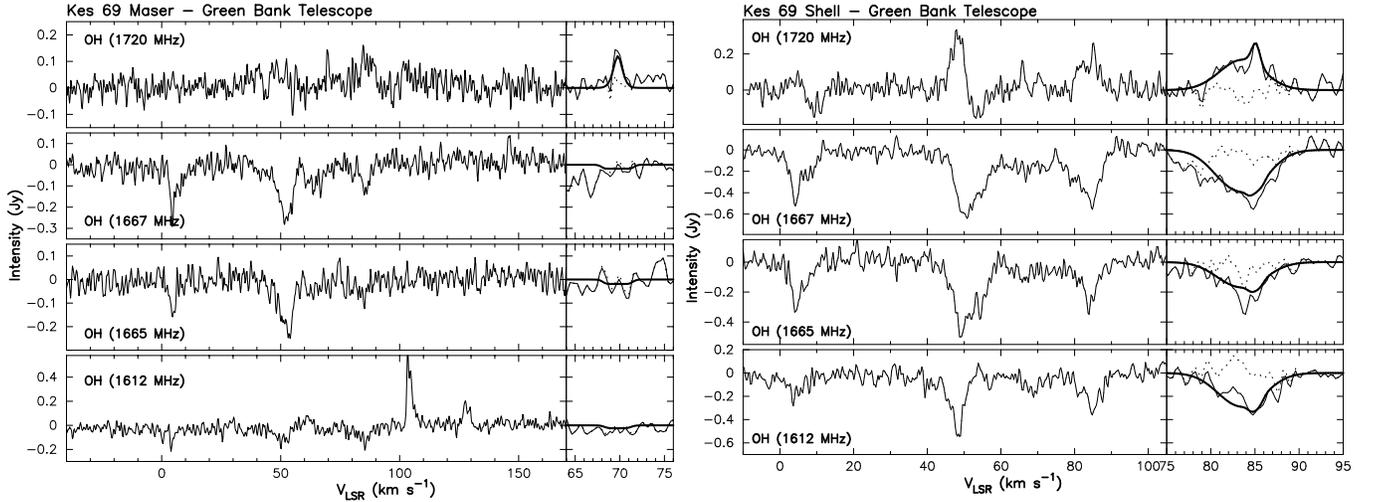}
\includegraphics[angle=-90,width=3.5in]{f5b}
\caption{Spectra of Kes 69 as in Figure 3.
The spectra towards the Kes 69 maser which previously detected at +69 \kms\ also shows the 1612 MHz profile of an OH/IR star in emission centered at 110 \kms . Spectra towards the bright southern shell of Kes 69 where 1720 MHz emission near +85 \kms\ shows a complex profile perhaps due to a superposition of maser emission within this velocity range.
In both figures Galactic clouds are seen in OH absorption and anomalous emission at +5 and +50 \kms . \label{fig:spec-kes69}}
\end{figure}

In addition to the compact maser, multiple OH spectral components are seen towards the bright radio shell of Kes 69. Absorption features at 1667/5 MHz and 1612 MHz is seen at velocities of +5, +50, +70 and +85\kms\ across the entire remnant. The spectra towards the compact maser in Figure $\ref{fig:spec-kes69}$ shows little absorption at 1720 MHz, but faint emission is present at $\sim$+85\kms . The flux of the maser matches that reported from VLA observations within our measured errors indicating no extended maser emission is associated with this compact maser. The fitted parameters given in Table $\ref{tbl:modelfits}$ are a high column and low filling factor consistent with compact maser emission.

\begin{figure} \centering
\includegraphics[angle=0,width=5.in]{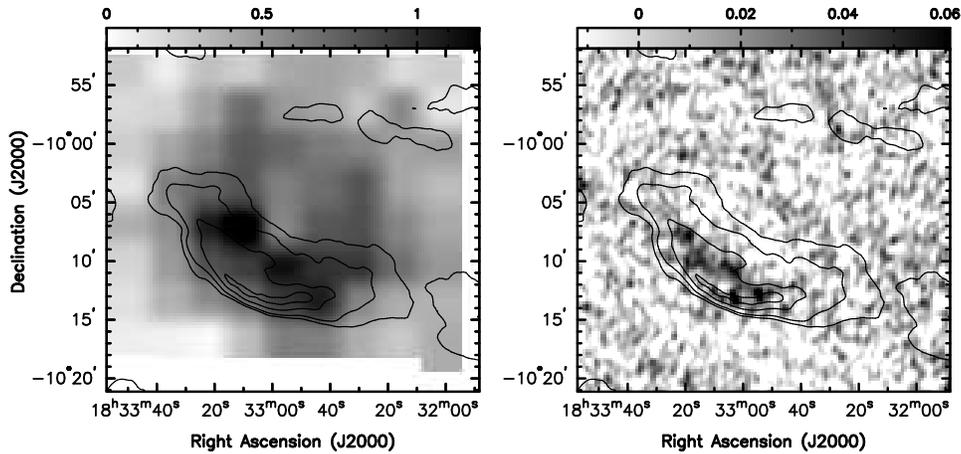}
\caption{[Left] Integrated line intensity of 1720 MHz maser emission towards Kes 69 observed with the GBT near +85 \kms\ in units of Jy \kms . [Right] Integrated line intensity of 1720 MHz maser emission towards Kes 69 near +85 \kms\ in units of Jy beam$^{-1}$ \kms . Both figures have 20cm radio contours overlaid at levels of 0.05, 0.1, 0.3 and 0.5 Jy \beam (synthesized beam of 45\arcsec ).
\label{fig:mom-kes69}}
\end{figure}

Towards the radio shell in Figure \ref{fig:spec-kes69} 1720 MHz emission is seen at +5 and +48\kms\ features with line widths of a few \kms , likely cases of anomalously emitting clouds present along the line of sight. The +85\kms\ feature also appears in emission across the southern ridge, but with broader line widths and a velocity gradient with velocity decreasing with increasing right ascension and declination. 
The integrated line intensity of emission near +85\kms\ is presented in Figure \ref{fig:mom-kes69}. 
We reanalyzed archival VLA observations\footnote{Project AY138 consisted of VLA D-array observations centered at 70\kms\ with a 0.2 MHz bandwidth.}
at 1720 and 1667 MHz to determine whether any low-level emission or absorption is detected towards the remnant. Because these observations were centered at +70\kms\ the +85\kms\ features fall at the edge of the spectral bandpass, but with careful calibration are apparent. Further VLA observations were obtained and will be presented in a future work clarifying the newly detected 1720 MHz emission. 

If the +85\kms\ emission is associated with the SNR as suggested here then the kinematic distance previously derived needs to be revised. The rotation curve of \citep{fich89} gives a tangent point at 7.9 kpc and +136\kms\ towards Kes 69. As was argued by \citep{green97}, if the maser is associated with Kes 69 and gives a systemic velocity for the remnant, then the presence of higher-velocity OH and H$_2$CO absorption features toward the remnant resolves the near/far distance ambiguity in favor of the far distance (11.2 kpc). However, if the systemic velocity matches that of the newly detected maser emission, the near distance is not ruled out. Recent HI and $^{13}$CO observations also show the highest absorption velocity towards Kes 69 being at +86\kms\ and a kinematic distance of 5.2 kpc (Tian Wenwu, private communication\footnote{The revised distance to Kes 69 was first presented by \citep{leahy2007aas}. The authors inform us that an error appears in the proceedings abstract. Kes 73 is incorrectly listed twice, and Kes 69 should appear in the first instance with a distance of 5.2 kpc.}). The detection of maser emission and molecular material at +85\kms\ casts some doubt as to whether the +69\kms\ maser is indeed associated with Kes 69. Wether both these disparate velocity components are associated with the same SNR remains an open question that needs to be resolved.

We note 1612 MHz maser emission from an OH/IR stars falls within the GBT beam in Figure $\ref{fig:spec-kes69}$ and is clearly seen with a center and outflow velocity of +115.6 and 11.4\kms , respectively. Another OH/IR star is also seen in our maps with v$_{LSR}$=+45.2\kms\ \citep{sevenster01}.

\subsection{G31.9+0.0 (3C391)}
The SNR 3C391 has a distinctive breakout morphology with a corresponding hydrogen column density gradient which rises towards the northwestern edge adjacent to a large molecular cloud \citep{chen04}. Further evidence of interaction comes from CO and CS line observations \citep{wilner98,reach99}. \citet{reach02} identified both molecular- and ionic-dominated regions, with a wide range of pre-shock densities and shock types being present.

\begin{figure} \centering
\includegraphics[angle=-90,width=3.5in]{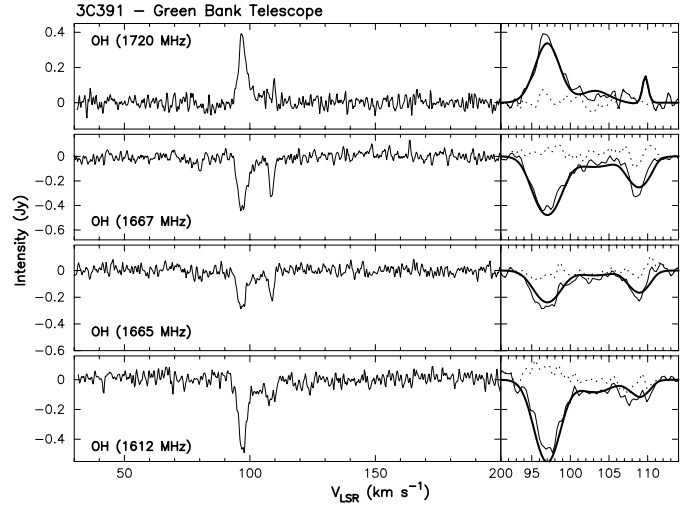}
\caption{Spectra of all four OH transitions towards 3C391 as in Figure \ref{fig:spec-w28}. \label{fig:spec-3c391}}
\end{figure}

\begin{figure} \centering
\includegraphics[angle=0,width=3.8in]{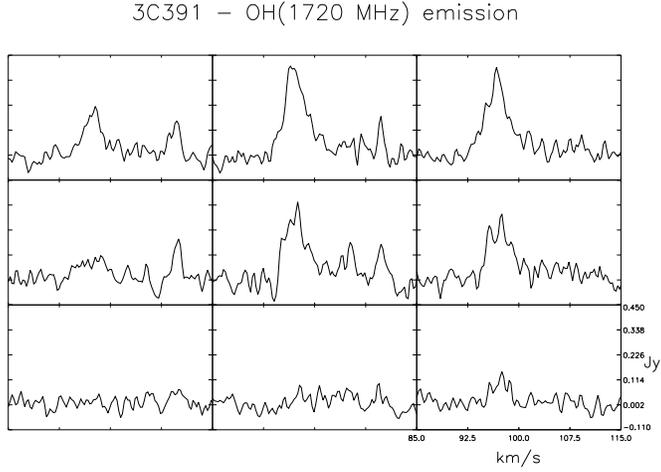}
\caption{Postage stamp plot of OH(1720 MHz) emission observed towards 3C391. Maser emission can be seen at +105 and +110\kms . The strongest 1720 MHz emission is associated with the with the SNR-cloud interaction along the bright northwestern ridge of the radio continuum at a velocity of $\sim$95\kms .\label{fig:post-3c391}}
\end{figure}

\citet{frail96} first identified OH masers associated with dense clumps along the northern and southern edges. Emission at $\sim$105\kms\ is associated with the broad molecular line (BML) region, while emission at $\sim$110\kms\ is associated with the interaction along the northeastern edge. We detect flux enhancements by factors of a few for both masers, suggesting additional masing gas is associated with these interaction regions. Spectra towards the BML and NE masing clumps as well as for the bright NW shell of the remnant have been modeled, with parameters given in Table $\ref{tbl:modelfits}$. Modeled spectra towards the peak in OH(1720 MHz) brightness are presented in Figure $\ref{fig:spec-3c391}$. 
Absorption is seen for OH main lines at +7, +23, +79 and +97 \kms\ in addition to absorption associated with the masers. The spatial distribution of the three OH velocity components associated with the remnant's vicinity are presented in Figure $\ref{fig:post-3c391}$. All lines show the strongest absorption at the center of the map where the continuum from 3C391 is the brightest. Inversion of the 1720 MHz line is seen at +7\kms\ across the map, likely an example of Turner's anomalous clouds. The strongest 1720 MHz emission is seen associated with the +97\kms\ feature to the north and west of 3C391 associated with the giant molecular cloud. It is unclear if this somewhat broad emission($\Delta$v = 3.8\kms ) is associated with low-gain extended masing post-shock gas, or anomalous emission associated with the cloud. No compact masers have been detected in this region.

\subsection{G34.7-0.4 (W44)}
One of the earliest identified interacting remnants, W44 has been extensively studied as a prototypical maser-emitting SNR. HI absorption indicates a kinematic distance of 2.8 kpc \citep{caswell75}. High-velocity CO(J=1-0) line wings are observed along the interaction boundary at densities of $\ga$10$^5$ cm$^{-3}$ \citep{seta04}. Infrared observations by ISO and Spitzer show strong ionic and molecular lines, requiring interaction with both a moderate and high density medium \citep{reach00,reach06}. Recently, \citet{castelletti07} identified a break in the spectral index at frequencies below 300 MHz. These regions of flat spectral index are argued to result from strong post-shock densities and enhanced magnetic fields.

OH(1720 MHz) emission was first reported toward W44 by \citet{goss71}. High-resolution observations established brightness temperatures in excess of 10$^6$ K and magnetic field strengths of $\sim$0.5 mG \citep{claussen97,claussen99,hoffman05_w44}. There are six clusters of detected masers denoted A through F shown in Figure $\ref{fig:mom-w44}$. With the GBT, significant enhancements of 1720 MHz line emission are detected in all maser groups, suggesting extended maser emission is present throughout the interaction region. 

Modeled spectra of these regions are presented in Figure $\ref{fig:spec-w44}$ and include several masing and non-masing components. For all spectra there is OH seen in absorption corresponding to the maser velocities, but there is also  a broadened absorption component at +40\kms\ that shows no associated 1720 MHz maser emission. The separation of the two components is most clearly seen in the spectra for clumps B and C. We also note there is clear evidence for a shock from the broad blue-shifted wing in OH absorption towards W44 with line widths extending up to 40\kms\ toward maser groups E and F. Clearly we are observing OH in the shocked gas accelerated both perpendicular to (in the case of masing gas) and toward the line of sight (in the case of non-masing gas with broad wings) within one 7\arcmin\ GBT beam.
Modeled parameters for the masing gas have OH columns $\ga$10$^{16}$ cm$^{-2}$ and larger filling factors than are seen for other remnants. The non-masing gas seen to peak in absorption at +40\kms\ is best fit by lower OH columns $\sim$10$^{16}$ cm$^{-2}$, systematically lower gas temperatures  and filling factors an order of magnitude higher than the masing gas. 

In addition to OH gas associated with the interaction, at +13.5\kms\ we detect narrow absorption at 1720, 1667 and 1665 MHz with emission at 1612 MHz indicative of a molecular cloud with anomalous 1612 MHz emission. This component is seen across the face of W44, but not outside the remnant.

\subsection{G49.2-0.7 (W51C)}
The complex region W51 is composed of two bright star forming regions W51A and B and the large supernova remnant W51C \citep{kundu67}. Observations at 151 MHz show W51C as a 30\arcmin\ shell which appears to connect to W51B along its western edge \citep{copetti91}. A low-frequency turnover of the radio continuum spectrum of W51C led to the suggestion that the SNR is surrounded or obstructed by a low-density ionized envelope \citep{copetti91}

While \citet{frail96} first searched the SNR for masers, the first identified detections were not reported until \citet{green97}. Shocked CO and HCO$^+$ indicate an interaction along the western edge of the remnant where the two bright masers are located \citep{koo97b}. Thus an association with the remnant is favored over a possible association with nearby star forming region W51B.
Masers associated with W51C have line-of-sight magnetic fields of 1.5 and 1.9 mG \citep{brogan00}.

Our GBT observations detect emission features at +69.2 and +72.1 \kms\ consistent with previous VLA observations of the SNR. The spectra showing two bright maser peaks is presented in Figure $\ref{fig:spec-w51c}$.  We find high OH column but low filling factors for both masers in our modeling. No evidence of extended maser emission is seen toward this remnant, though an extended interaction is evident from CO data. Furthermore, a shocked ridge 8$\arcmin$ in length is seen in 2MASS K$_S$-band emission surrounding the masers. The lack of extended maser emission may be due to unfavorable conditions for the production of OH: a fast, dissociative J-type shock destroys molecules and nearly all of the hydrogen in the shocked gas is found to be atomic, not molecular \citep{koo97b}. However, we do find a strong absorption component from the maser velocities to +60 \kms . This absorption appears to be related to the shock interaction as it is only present toward the shocked arc and is not correlated with the background continuum. An anomalous 1720 MHz cloud is seen along the western edge of the map at +5.5\kms .

\subsection{G189.2+3.0 (IC443)}
The SNR IC443 (G189.2+3.0) is one of the best sources in which to study the physical processes of a supernova remnant interacting with adjacent molecular clouds. Located at a distance of 1.5 kpc in a direction largely free of the confusion present toward the Galactic plane, it was well suited for some of the earliest studies of molecular tracers of shock excitation (DeNoyer 1979; Huang, Dickman \& Snell 1986; Burton 1987; Tauber et al. 1994). Evidence of IC443's interaction was first observed as large line width, negative velocity neutral hydrogen by DeNoyer (1978). Huang, Dickman \& Snell (1986) identified a series of shocked CO clumps (designated A-H) along the southern ridge, with sizes ranging from 0.5 to 5 pc. Further high-resolution study revealed shocked molecular line emission tracing out a ring of perturbed molecular gas about the center of the remnant (Dickman et al. 1992; van Dishoeck, Jansen \& Phillips 1993).

Initial surveys for OH(1720 MHz) masers detected only the bright maser group G along the western edge of the shocked molecular ridge. With the GBT we were able to identify two new clumps (B and D) of maser emission along the south-eastern ridge \citep{hewitt06}. These new masers are detected with sufficient flux density that they should have been recognized by earlier VLA observations. We therefore concluded that masers B and D were either extended maser emission regions which were resolved out by the VLA, or are composed of a cluster of fainter maser spots that lies below detectable limits of previous observations. This demonstrated the ability of single dish observations to detect low-level and extended maser emission even in the absence of strong compact masers. 
%%% NO FIGURES?
%%% IS INCLUDED IN MODEL FITS... need to mention they are updated in Table 2.
Further details can be found in \citet{hewitt06}. 

\subsection{G348.5+0.1 (CTB 37A) and G348.5--0.0}
CTB 37A is a bright remnant with a distinctive breakout morphology along its southwestern extent. It is thought that a second, fainter shell-type SNR overlaps the shell of CTB 37A along it's northern extent. 
The SNR G348.5--0.0 is seen only as a partial shell overlapping with CTB 37A \citep{kassim91}, though it is clear from both detected OH(1720 MHz) masers \citep{frail96} and associated CO clouds \citep{reynoso00} that each are distinct SNRs with systemic velocities of --65\kms\ and --22\kms\ respectively.
Hereafter we consider the two SNRs CTB 37A and G348.5--0.0 separately, though 
our 21\arcmin\ by 14\arcmin\ map and the spectra presented in Figure $\ref{fig:spec-ctb37a}$ include both SNRs with distinguishable velocity components. 

Towards CTB 37A \citet{frail96} detected OH(1720 MHz) masers between --63 and --67\kms . Three clouds associated with the interaction are observed in CO near --65\kms\ \citep{reynoso00}. The northernmost cloud appears to lie coincident with infrared source IRAS 17111-3824, which has colors indicative of shock-heated dust \citep{reynoso00}. Both SNRs are detected in the infrared with Spitzer \citep{reach06}. Infrared colors indicative of molecular shocks are seen toward the northern-most OH(1720 MHz) masers. Figure $\ref{fig:post-ctb37a}$ gives a postage stamp plot of the --65\kms\ maser emission associated with CTB 37A. The integrated maser-line intensity is consistent with previous VLA measurements suggesting there is little or no maser emission on spatially extended scales. The GBT spectrum toward the bright --63.5\kms\ masers is presented in Figure $\ref{fig:spec-ctb37a}$, with some tentative evidence for shallow, wide absorption at the other main-line transitions.The parameters of best-fit are for an OH column of $\sim$10$^{16}$ cm$^{-2}$ and very small filling factors consistent with compact, dense maser clumps.

Toward G348.5--0.0 two masers were detected at --21.4 and --23.3\kms\ , with GBT observations recovering the same 1720 MHz flux as previous VLA observations. The observed OH spectra show corresponding OH absorption in the mainlines centered at --22\kms\ with a FWHM of $\sim$5\kms . A cloud was detected in CO observations \citep{reynoso00}.
Figure $\ref{fig:post-g3485}$ gives a postage stamp plot showing the distribution of the --22\kms\ component, with a spectra of G348.5--0.0 given in Figure $\ref{fig:spec-ctb37a}$. Model parameters indicate that the masing clump has an order of magnitude higher OH column and order of magnitude smaller filling factor when compared to the OH gas associated with this --22\kms\ absorption.

We also note several anomalous 1720 MHz emission components with corresponding absorption are present towards CTB 37A and G348.5--0.0. To the north and west covering most of our map is a --9\kms\ cloud. We are able to easily fit the observed OH profiles using the same model-fitting but with a density of 100 cm$^{-3}$ instead of 10$^5$ cm$^{-3}$ typical of masers. The best-fit filling factor is consistent with a spatially extended cloud. At the northwest edge of our map are additional anomalous 1720 MHz clouds at --102.3\kms\ and --112.9\kms . These western features appear to be associated with the 3 kpc arm of our galaxy, consistent with previous HI and CO observations which place both remnants at the far kinematic distances of $\ga$11.3 kpc \citep{caswell75,reynoso00}. 

\subsection{G349.7+0.2}
At a kinematic distance of 22.4 kpc, G349.7+0.2 is the most distant and luminous SNR within our galaxy with detected OH(1720 MHz) masers. The shock interaction is indicated by numerous molecular transitions including $^{13}$CO, $^{12}$CO, CS, H$_2$, HCO$^+$, HCN, H$_2$CO and SO \citep{lazendic04oh}. The cloud being struck by the expanding SNR is identified in CO observations at $v_{\rm LSR}$=+16.2\kms\ with a linear size of about 7 pc, a mass of $\sim $104 $M_\odot$ and a volume density of $\sim $10$^3$ cm$^{-3}$ \citep{dubner04}. The shock interaction is thought to be taking place primarily on the far side of the SNR from a comparison between total absorbing column density and slightly red-shifted wings in the optically thin $^{13}$CO spectra. It is a member of the mixed morphology class of remnants having a central thermal X-ray plasma \citep{lazendic04}.

\citet{frail96} identified five maser spots at velocities between +14.3 and +16.9\kms . Zeeman splitting is observed for only one of the three bright masers above 200 mJy \citep{brogan00}. The +15\kms\ maser has a line-of-sight magnetic field strength of 0.35 mG, while other bright masers have comparable non-detection limits. Our GBT observations detect an integrated line intensity of 3.53 Jy \kms , a marked increase from the 2.15 Jy \kms\ resolved by the VLA in A configuration \citep{brogan00}. This suggests that despite its small diameter, significant emission at lower brightness temperatures is present across the remnant. Due to the bright continuum we did not achieve sufficient sensitivity to resolve any corresponding main-line absorption. Our models are thus somewhat poorly constrained yielding an OH column of 10$^{17}$ cm$^{-2}$ and filling factor of 0.005, which appears consistent with other remnants. The spectrum is presented in Figure $\ref{fig:spec-g3497}$. Absorption components are seen at --110, --64 and --20 \kms\ with anomalous 1720 MHz clouds identified at --95 and --61 \kms . We are able to produce reasonable fits to the lines using our models at densities of 10--1000 cm$^{-3}$ and low temperatures, in agreement with Galactic clouds along the line of sight. 

\subsection{G357.7-0.1 (Tornado)}

The Tornado Nebula is so named for its distinctive elongated radio continuum structure. It has a non-thermal spectral index consistent with it being a shell-type SNR and a central thermal X-ray plasma placing it in the mixed morphology class of supernova remnants \citep{fyz03mmsnrs}.
Chandra observations reveal X-ray emission with a head-tail structure \citep{gaensler03} with the interior thermal plasma at a temperature kT$\sim$0.6 keV. A distance of 12 kpc is derived from neutral hydrogen column N$_H\sim$10$^{23}$ cm$^{-2}$. The eye of the Tornado (G357.63-0.06) appears to be an unrelated thermal source. H92$\alpha$ recombination line emission present at {\rm--210}\kms\ and \hi\ absorption indicate the \hii\ region is located at a distance near the Galactic center \citet{burton04}. Polarized loop structures are also seen as the tail, which are thought to be a spiral magnetic field structure in which the transverse magnetic field is tangential to these loops \citep{stewart94}.

Multiple OH(1720 MHz) masers are detected at the western edge of the remnant near the head of the Tornado \citep{frail96}. Polarization measurements determine a line-of-sight magnetic field of +0.7$\pm$0.12 mG for the brightest maser \citep{brogan00}.  It is clear from molecular observations that a shock interaction occurs where the OH(1720 MHz) masers are observed \citep{lazendic04}.
With short VLA D-array observations lower level maser emission is seen to the south of the head and along the eastwardly extending tail \citep[Fig. 4]{fyz99}.

A grid of four by two GBT pointings cover the extent of the Tornado Nebula. We note that interpreting the OH(1612 MHz) spectrum for this source is difficult due to the presence of three OH/IR stars  in our small map \citep{sevenster97}. Source OH357.675-0.060 dominates the 1612 MHz spectrum and offset at a central velocity of -237.4\kms\ such that our frequency shifting causes the 1612 MHz emission to be folded over the absorption. For these scans we did not use folding and instead examined the uncontaminated portion of the 1612 MHz spectrum.

The GBT recovers five times more maser-line flux than previously detected in compact maser sources with the VLA \citep{fyz99}. The fitted spectrum is presented in Figure $\ref{fig:spec-tornado}$ with bright masers seen between --15 and --12.0\kms . In our spectral modeling of the best-fit filling factors are more than an order of magnitude higher than those found for compact maser spots, indicating most of the 1720 MHz line flux is on large scales. 
Additionally we see an anomalous 1720 MHz cloud at +4.5\kms\ across the western half of our map. OH absorption components are also clearly seen at +23, --4, --38 and --60 \kms .

\subsection{G357.7+0.3 (Square)}
The Square nebula is a supernova remnant with a distinctive square morphology \citep{gray94}. Despite being a classic case of a maser-emitting SNR it is perhaps the poorest studied remnant at wavelengths outside the radio spectrum.
We observed a 40' by 47' region around the full SNR extent, and then remapped a 20' by 24' (5 by 6 pointings) region around the bright maser peak for improved sensitivity. 
As noted by \citet{fyz99} OH(1720 MHz) emission is large in extent, extending up to 20$\arcmin$. A strong, broad emission region appears to arise from a superposition of narrower velocity components centered at --35\kms\ as is seen in the integrated line intensity map in Figure $\ref{fig:mom-square}$. The maser emission extending across the western edge also shows a velocity gradient of a few \kms .

The spectra towards compact maser source A is presented in Figure $\ref{fig:spec-square}$. Detected 1720 MHz emission centered at --35\kms\ appears to be somewhat broad, likely a result of the velocity gradient across our GBT beam. Strong symmetric absorption is seen for the other ground-state lines but broadened up to some 40\kms , which is another indicator that the SNR is interacting with a nearby cloud. The symmetry of the absorption indicates the shock is propagating largely in a direction tangential to the line of sight, a geometry which is most favorable to maser amplification. We see a ten-fold increase in the total 1720 MHz flux as compared to the VLA. The modeled spectra is best-fit by OH columns near 10$^{17}$ cm$^{-2}$ and large filling factors for the maser emission. Clearly a large extended maser structure is present along the full cloud-shock boundary at the western edge of the Square.

An anomalous 1612 MHz cloud extends across our map and appears to have a small velocity gradient from +2\kms\ in the easternmost edge to -2\kms\ in the westernmost edge of our map. The distribution of this cloud is shown in Figure $\ref{fig:mom-square}$(bottom right). Additionally there is a cloud at +10\kms\ in the center of our map which is only visible in OH main-line absorption.

\subsection{G359.1--0.5}
The supernova remnant G359.1--0.5 is a mixed-morphology remnant which shows center-filled X-rays with prominent K$\alpha$ lines \citep{bamba00}. Five regions of compact masers are detected throughout the radio continuum shell \citep{fyz96}. Located near the Galactic center in projection, a bright non-thermal radio filament crosses the bright radio continuum shell along the northwestern edge. There appears to be no connection between the two structures, but the presence of OH(1720 MHz) masers at the point where the two radio continuum structures overlap led to detailed studies of the shock interaction. A bar of shocked molecular hydrogen with broad line profiles seen for CO, HCO+ and CS were observed towards the OH(1720 MHz) masers at this position \citep{lazendic02}. However, recent XMM observations indicate that the distance to G359.1--0.5 is not consistent with it being located at the Galactic center, but at a distance of $\sim$5 kpc and possibly associated with a nearby pulsar wind nebulae \citep{fyz07}.

We observed a 17\arcmin\ by 31\arcmin\ region covering the western half of G359.1--0.5. In this map three of the five maser groups are visible, including maser group A which was previously studied using near-IR and millimeter lines. Figure $\ref{fig:mom-g3591}$ shows a map of the integrated 1720 MHz line intensity, with masers seen along the entire extent of the radio shell covered by our map. Comparison with previous VLA observations indicate that the flux of masers in G359.1--0.5 is significantly higher in our GBT measurements, consistent with extended masing regions.

The spectrum toward maser A is presented in Figure $\ref{fig:spec-g3591a}$. Maser emission is seen at --5\kms\ and broad absorption centered at --11\kms\ shows a somewhat blue-shifted asymmetry. Absorption is remarkably broad, with a full width at half-maximum at $\sim$50\kms , indicative of shock accelerate gas.
Additionally we note that OH absorption features are seen at +6 and +16\kms\ but only in the northern extent of the SNR where there is some confusion with other radio continuum features.

\subsection{Galactic center remnants: G0.0+0.0 (Sgr A East), G1.0-0.1 (Sgr D SNR) and G1.4-0.1}
For these three SNRs, deep broad absorption from foreground clouds greatly hinders analysis of maser-line intensities. Figure $\ref{fig:spec-sgra}$ shows the 1720 MHz spectrum from --200 to +200\kms\ towards the bright 10 Jy maser in Sgr A East. While maser emission is seen at +66\kms\ it is clearly affected by the --20 Jy absorption dip in the spectrum. A postage stamp plot of the entire mapped region is shown in Figure \ref{fig:post-sgra}.

\section{Discussion}

\subsection{Extended OH(1720 MHz) Emission} \label{sec:eme}

It is clear from our observations that significant flux from 1720 MHz masers is missed by previous observations. A comparison of integrated maser line intensity between GBT and published VLA measurements is given in Table $\ref{tbl:eme}$. We give the integrated line intensity\footnote{The integrated line intensity $\gamma = S_{p} \frac{2 (\Delta {\rm v})}{\sqrt{\pi/ln(2)}}$ is measured in units of Jy \kms\ where S$_{p}$ is the peak flux density and $\Delta$v is the full-width at half maximum as fit by a Gaussian. This is also commonly referred to as the $"$zero-eth moment$"$ of a Gaussian.} as it is a robust measure of maser emission less sensitive to channel width if the narrow maser line is not fully sampled. Absolute flux calibration for the GBT is estimated to be better than 10$\%$. Enhancements significantly exceeding this uncertainty are interpreted as evidence for the presence of low gain or extended maser emission which is resolved out due to the interferometer's lack of sensitivity to low brightness temperatures and large spatial scales. 

For ten of the thirteen remnants considered here we find a significant maser flux enhancement. Thus extended OH(1720 MHz) maser emission appears to be a common phenomenon.
The total maser flux is seen to generally increase as the angular size of the remnant increases and as the distance to the source decreases.
For larger, closer remnants one might expect to find extended maser emission present on larger scales, and thus the VLA would be less sensitive to an increasingly higher portion of the extended maser emission. However, such a correlation between the GBT/VLA flux ratio and the physical or angular size of the remnant is not seen in our data. As maser gain is highly non-linear, bright compact spots may or may not outshine the weaker large scale extended flux we observe only with the GBT, producing a natural scatter in the GBT/VLA ratio. 

Bright maser emission over several GBT pointings is seen for W28, W44, the Square and G359.1--0.5 in Figures \ref{fig:mom-w28}, \ref{fig:mom-w44}, \ref{fig:mom-kes69}, \ref{fig:mom-square} and \ref{fig:mom-g3591}, respectively. However, compact maser spots are distributed throughout the observed areas of emission. We have attempted to isolate the extended emission by subtracting a model of compact masers detected with the VLA from our GBT maps. There is no clear difference in morphology between the model subtracted and total integrated emission maps. Though we are not able to determine much from 

Interestingly, for remnants IC443 and Kes 69 we detect spatially distinct regions of maser emission that were not detected by previous VLA observations. These new masers are observed with brightnesses some twenty times lower than corresponding compact masers but with narrow line widths($\sim$ 2 \kms ) and low brightness temperatures(T$_B$ $\le$ 2500 K) consistent with the properties of extended maser emission. In our modeling of IC443 we find extended masers B and D have lower columns, higher filling factors and lower excitation temperatures in comparison to compact maser G. The new masers are also located in regions in which a SNR shock has been identified from broad molecular lines and suggest that extended maser emission more fully traces the extent of interaction.

Further observations are needed to resolve these newly identified maser regions, but for a number of remnants VLA observations in the D configuration have already detected some evidence of extended low level structures. Short observations of W28, W44, the Tornado, the Square and G359.1-0.5 all indicate the presence of extended maser structures tracing the large-scale shock interaction \citep{fyz95,fyz99,fyz03_w28,hewitt07}. Extended maser emission appears to be common for maser-emitting supernova remnants, and has the potential to be a very useful shock diagnostic on large scales, with all the diagnostic benefits of compact maser emission.

\subsection{Modeling \label{sec:modeling}}
From theoretical considerations, an inversion of the OH 1720 MHz line behind shocks is only maintained for a strict range of physical conditions: moderate temperatures of 50Ð-125 K, densities of 10$^5$ cm$^{-3}$ and OH column densities of the order 10$^{16}$ to 10$^{17}$ cm$^{-2}$ \citep{lockett99}. Observations of all four ground-state transitions with the GBT permits us to estimate the parameters of the gas responsible for producing the observed line profiles using a simple multi-component model, as we have successfully demonstrated for SNR IC443 \citep{hewitt06}.

Spatial blending of compact maser components leads to inherent uncertainties in our single dish data. We do not perform a detailed model fit to the line profiles (which in any case are not Gaussian); instead we have attempted to fit each spectra with the minimum number of emission and absorption components such that a reasonable fit is found. The derived parameters for SNRs are presented in Table \ref{tbl:modelfits} and give a good illustration of the physical conditions of both maser emission (where T$_{ex}^{1720}<$ 0 K) and post-shock absorbing gas from the observed remnants. 
Generally, a lower OH column is associated with a higher filling factor indicating a more extended emission region. Maser emission is seen for a broad range of filling factors, most often fitted with low filling factors indicating most of the emission originates from smaller regions. 

We also attempt to fit a few clear instances of anomalous OH emission from clouds in which the SNR serves as a bright background continuum source. These fits are listed separately in Table \ref{tbl:anomfits}. Anomalous OH(1720 MHz) emission is generally seen for larger filling factors, low temperatures and modest OH column densities, as expected if arising in lower density Galactic clouds, and is discussed further in Section $\ref{sec:anom}$.

Plots of OH column versus filling factor, OH column versus kinetic temperature and filling factor versus line width are given in Figure 22. Trends in the physical conditions can be seen where maser-emitting regions tend to have higher OH column and kinetic temperature than corresponding broad absorption regions. Kinetic temperatures are systematically lower for the broad absorption components with little variance seen between multiple maser components in each remnant. 
The line-of-sight OH column $\noh$ for masers is found to lie between 8$\times$10$^{15}$ and 1.6$\times$10$^{17}$ cm$^{-2}$. For the derived molecular densities of 10$^5$ cm$^{-3}$ this roughly spans the range for which 1720 MHz can be inverted in our model (see Figure 2, Wardle 2007). Fitted columns for absorption profiles span a larger range of OH column, but similarly are not above $\sim$10$^{17}$ cm$^{-2}$. At a column of $\ga$10$^{17}$ cm$^{-2}$ the models of \citet{wardle07} predict that the OH(6049 MHz) line is also inverted. However, searches towards the remnants modeled in this work for OH transitions which are inverted at such higher column densities have all been negative \citep{mcdonnell07,pihlstrom08,fish08}. This suggests there may be an upper limit of $\noh \la$10$^{17}$ cm$^{-2}$ toward interacting supernova remnants. This limit is particularly noticeable for masers and SNR absorption profiles in Figure 22 but no such limit is seen for anomalous emission from clouds.

The physical properties we derive for extended maser emission appear consistent with the idea that it results from extended post-shock regions in which physical conditions and geometry and produce only a modest $\noh$. As can be seen in Figure 3 of \citet{lockett99} inversion of the 1720 MHz line is maintained over densities of 10$^3$ to a few times 10$^5$ cm$^{-3}$ with $\noh \ga$ 10$^{15}$ cm$^{-3}$. For extended structures where the OH column is low, only low-gain amplification occurs, resulting in low brightness temperatures that can be observed with more sensitive synthesis observations.

\subsubsection{Significance of the Filling Factor}
Model fitting uses a filling factor {\it f} to represent the fact that often emission and absorption components originate from a region which does not fill the GBT beam. From equation $\ref{eq:T_v}$ it can be seen that {\it f} is the beam filling factor under the assumption that the entire GBT continuum lies behind the emitting or absorbing region. For compact masers this is certainly not the case, and the appropriate spatial scale is a product of both the beam filling factor and a continuum filling factor which indicates the ratio of continuum flux which lies behind the emitting/absorbing region.

To verify that the filling factors introduced in our modeling are an accurate indicator of the relative spatial scales we examine the cases of W51C for which there is no excess maser emission detected by the GBT. All detected maser emission is due to point sources as detected by the VLA. The ratio between the area of these point sources and the GBT beam is found to be approximately equal to the fitted filling factor times the ratio of 20 cm continuum flux from behind the maser to the total flux within the GBT beam, as expected for compact sources.

For the maser emitting regions which were previously resolved out or not detected by the VLA we generally find larger filling factors. If this emission is present on very small spatial scales then almost none of the continuum detected with the GBT beam is behind the emitting regions. High resolution continuum maps exist for all of the supernova remnants considered here, and it is seen to be fairly well distributed across the shell. Thus it is highly unlikely that the vast majority of the continuum emission is not behind the emitting regions, and the larger filling factors are likely to be indicative of larger spatial scales. Future observations at higher resolution but still sensitive to large spatial scales will resolve the distribution of excess maser emission detected by the GBT and will remove this ambiguity in the modeling.

\subsection{Anomalous OH Clouds \label{sec:anom}}

It is important to differentiate extended maser emission from the widespread Galactic 1720 MHz emission detected by \citet{turner82} which is associated with giant clouds. Towards several SNRs we detect modest 1720 MHz inversions which appear somewhat too broad for maser emission and at a velocity distinct from the observed masers and broad absorption. Additionally, the emission shows a correlation with the background continuum. The so called "anomalous" OH emission cataloged by \citet{turner82} is characterized by $\sim$5 \kms\ line widths and an excellent correlation with spiral arms throughout the Galactic plane. To derive how the physical conditions of these anomalous emission regions differ from the observed maser emission we have used our excitation models of hydroxyl but at densities between 100 and 10$^{3.5}$ cm$^{-3}$ appropriate for molecular clouds. The resulting model fits for a few anomalous clouds are given in Table \ref{tbl:anomfits}.

Extended OH(1720 MHz) maser emission associated with interacting remnants can be differentiate from anomalous OH emission by several characteristics: Maser emission appears within the boundary of the SNR radio shell, but shows no correlation with the strength of the radio continuum. Instead a clear correlation with molecular shock tracers is seen. Narrow line widths of a few \kms\ or less are seen for both compact and extended features, much less than the line width of molecular clouds. The range of physical conditions for extended masers in SNRs are also quite different than those for anomalous 1720 MHz emission.
With sensitive interferometric observations extended OH(1720 MHz) maser emission can be resolved tracing the extension of the large scale interaction between the SNR and adjacent clouds.

Study of the giant cloud G28.17+0.05 shows that 1720 MHz emission arises only from within the cold cloud core where there is associated $^{12}$CO emission \citep{minter01}. Collisional pumping of OH at low temperatures (15-40 K) in the cloud core has been suggested as a mechanism to explain the anomalous 1720 MHz emission.
\citet{minter01} suggest that there are at least 100 large clouds analogous to G28.17+0.05 in the inner Galaxy, giving a good probability that any line of sight through within 50\degr\ of the Galactic center probably intersects at least one. Single-dish surveys for OH(1720 MHz) masers observed 141 SNRs, with 53 having detected OH(1720 MHz) emission, almost all of which lie within 50\degr of the Galactic center \citep{frail96,green97}. Synthesis follow-up found only 12 SNRs contained OH(1720 MHz) masers with the other single dish detections being attributed to anomalous emission from giant clouds.

In a few cases we detect clouds which show strong absorption at 1665, 1667 and 1720 MHz but emission at 1612 MHz. For these cases where we see a somewhat broad ($\sim$3\kms ) inversion of the 1612 MHz line further investigation of the excitation conditions is needed. \citet{elitzur76} found that as temperature is increased above $\sim$70 K collisional inversion of the 1612 MHz line and anti-inversion of the 1720 MHz line occurs for density ranges similar to those of 1720 MHz inversions (see their Figure 3). It is possible this indicates GMCs which have been collisionally heated above 70 K. Alternatively, there is a critical OH column above which 1720 MHz inversion is replaced by 1612 MHz inversion. Presently the best case of anomalous 1612 MHz inversion is the +7\kms\ cloud along the line of sight towards W44.

Though we only have identified a handful of anomalous OH clouds, it is worthwhile to examine their distribution in the Galaxy. Figure $\ref{fig:galdia}$ shows the distribution of masers and anomalous OH clouds in a Galactic longitude-velocity diagram. Comparison with the CO survey of Dame et al. (1987) shows that anomalous OH clouds all lie within the nuclear ring or molecular disk. As more anomalous clouds are identified it will be possible to better test the claim of \citet{turner82} that these clouds are giant molecular clouds tracing the spiral arms of our galaxy.

\section{Conclusions}

We have observed \numobssnr\ remnants in all four ground-states of OH with the GBT. For nine of the remnants we detect significantly more flux with the GBT than has been observed with previous interferometric observations indicating the presence of a significant amount of maser emission on extended scales.  Additionally we find new maser emitting regions for SNRs Kes 69 and IC 443, possibly indicating that extended maser emission is an accurate tracer of the shock interaction even in the absence of bright compact masers. For W28, G16.7+0.1, 3C391, W44, G349.7+0.2, the Tornado and the Square SNRs we find evidence of enhanced 1720 MHz emission, suggesting such extended maser emission is quite common for maser-emitting remnants.

Using OH excitation modeling we have produced good fits to the observed line profiles. The physical parameters of the absorbing and emitting gas observed at the SNR/cloud interaction are found to be consistent with theoretical expectations of OH(1720 MHz) masers forming in molecular gas with temperatures of 30--100 K, OH columns of 10$^{15-17}$ cm$^{-2}$ and densities of $\sim$10$^5$ cm$^{-3}$. Additionally we have identified several new instances of anomalous OH 1720 and 1612 MHz clouds as previously classified by \citet{turner82}. We find that lower densities and gas temperatures produce good fits via our OH excitation modeling for such clouds, and it is possible to differentiate them from extended OH(1720 MHz) masers.

Extended maser emission appears to be a common phenomenon which is a useful tracer of conditions in cooling gas just behind the shock front throughout the region in which a supernova remnant is interacting with an adjacent molecular cloud.

\begin{acknowledgments}
We thank Ron Maddalena, Jim Braatz and the GBT operators for their help throughout these observations. Support for this work was provided by the NSF through award GSSP06-0009 from the NRAO.
\end{acknowledgments}

\facility{ {\it Facilities:} \facility{GBT}, \facility{VLA}}

\bibliography{jackref}

\clearpage

\begin{table}
\centering
\caption{Observations of OH toward supernova remnants \label{obstbl}}
\begin{tabular}{rllrrr} \hline
SNR & Name & \multicolumn{2}{c}{Map Center} & Grid Size  & Map Extent \\
    &             & $\alpha_{J2000}$ & $\delta_{J2000}$ \\
\hline
 0.0+0.0   & Sgr A East& 17:46:00 & --28:55:56 &10x10 & 37$\arcmin$x37$\arcmin$ \\ 
 1.0--0.1  & Sgr D SNR & 17:48:44 & --28:08:55 & 3x6 & 14$\arcmin$x21$\arcmin$ \\
 1.4--0.1  &         & 17:49:34 & --27:48:13 & 3x3 & 14$\arcmin$x14$\arcmin$ \\
 6.4--0.1  & W28     & 18:00:56 & --23:26:03 & 18x12 & 63$\arcmin$x44$\arcmin$ \\
 16.7+0.1  &         & 18:20:58 & --14:20:30 & 3x3 & 14$\arcmin$x14$\arcmin$ \\
 21.8--0.6 & Kes 69  & 18:32:56 & --10:05:15 & 9x8 & 34$\arcmin$x31$\arcmin$ \\
 31.9+0.0  & 3C391   & 18:49:25 & --00:56:30 & 3x3 & 14$\arcmin$x14$\arcmin$ \\
 34.7--0.4 & W44     & 18:56:02 & +01:19:57 & 9x10 & 34$\arcmin$x37$\arcmin$\\
 49.2--0.7 & W51C    & 19:23:18 & +14:10:52 & 10x6 & 37$\arcmin$x21$\arcmin$\\
189.1+3.0  & IC 443  & 06:16:56 & +22:29:08 &14x13 & 51$\arcmin$x48$\arcmin$ \\
348.5+0.1  & CTB 37A & 17:14:03 & --38:36:28 & 6x3 & 21$\arcmin$x14$\arcmin$ \\
349.7+0.2  &         & 17:18:00 & --37:26:00 & 2x2 & 11$\arcmin$x11$\arcmin$ \\
357.7--0.1 & Tornado & 17:40:30 & --30.56.32 & 4x2 & 17$\arcmin$x11$\arcmin$ \\
357.7+0.3  & Square  & 17:39:14 & --30:38:40 & 11x13 & 40$\arcmin$x47$\arcmin$\\
359.1--0.1  &         & 17:44:53 & --29:56:13 & 4x8 & 17$\arcmin$x31$\arcmin$\\
\hline
\end{tabular}
\bigskip
\end{table}

\begin{table}
\centering
\caption{Best-fit Model Parameters \label{tbl:modelfits}}
\begin{tabular}{lrrrrccccccccccccccccc}
\hline \hline
{SNR} &
{$v_0$} & 
{$\Delta v$} & 
{log $\noh$} & 
{$f$} & 
{log $n$} & 
{T$_k$} & 
{$\tex^{1612}$} &  
{$\tex^{1665}$} &  
{$\tex^{1667}$} & 
{$\tex^{1720}$} \\
Name & {(\kms )} & {(\kms )} &{ $({\rm cm}^{-2})$} & & {(cm$^{-3}$)} & {(K)} & {(K)} & {(K)} & {(K)} & (K) \\
\hline
W28 A
& +6.45 $\pm\ $0.02& 1.15 $\pm\ $0.04& 16.93 $\pm\ $0.14& 0.0067 $\pm\ $0.0007& 5.0 & 75 & 1.78 & 4.44 & 4.14 & --13.55\\
& +5.76 $\pm\ $0.02& 4.70 $\pm\ $0.02& 17.14 $\pm\ $0.01& 0.0021 $\pm\ $0.0007& 5.0 & 75 & 2.40 & 5.41 & 4.71 & --3.92\\
& +2.45 $\pm\ $0.38& 29.1 $\pm\ $0.9 & 16.39 $\pm\ $0.06& 0.034 $\pm\ $0.0009 & 5.0 & 60 & 4.61 & 4.88 & 4.99 & 5.28\\
W28 E,F
& +10.01 $\pm\ $0.01& 1.89 $\pm\ $0.01& 17.14 $\pm\ $0.01& 0.0006 $\pm\ $0.0001& 5.0 & 75 & 2.39 & 5.41 & 4.71 & --5.28\\
& +11.41 $\pm\ $0.01& 2.67 $\pm\ $0.01& 17.14 $\pm\ $0.01& 0.0006 $\pm\ $0.0001& 5.0 & 75 & 2.41 & 5.41 & 4.71 & --3.50\\
& +15.15 $\pm\ $0.01& 3.11 $\pm\ $0.01& 16.03 $\pm\ $0.01& 0.0028 $\pm\ $0.0004& 5.0 & 75 & 0.96 & 4.16 & 3.07 & --2.32\\
& +13.90 $\pm\ $0.02& 4.05 $\pm\ $0.01& 17.14 $\pm\ $0.01& 0.0005 $\pm\ $0.0001& 5.0 & 75 & 2.41 & 5.41 & 4.71 & --3.11\\
&  +8.16 $\pm\ $0.01& 7.88 $\pm\ $0.02& 17.14 $\pm\ $0.01& 0.0254 $\pm\ $0.0004& 5.0 & 75 & 2.39 & 5.41 & 4.71 & --5.21\\
&  +7.35 $\pm\ $0.02& 5.99 $\pm\ $0.04& 16.34 $\pm\ $0.05& 0.043 $\pm\ $0.001  & 5.0 & 50 & 6.87 & 7.33 & 7.59 & 8.11\\
&  +7.08 $\pm\ $0.04& 20.4 $\pm\ $0.1 & 16.40 $\pm\ $0.05& 0.038 $\pm\ $0.001  & 5.0 & 60 & 4.62 & 4.87 & 5.00 & 5.28\\
&  +7.32 $\pm\ $0.03& 41.7 $\pm\ $0.1 & 16.15 $\pm\ $0.02& 0.184 $\pm\ $0.001  & 5.0 & 50 & 6.80 & 7.29 & 7.55 & 8.10\\
\hline
G16.7+0.1
 & +19.92 $\pm\ $0.11& 1.87 $\pm\ $0.16& 16.9 $\pm\ $0.6& 0.0051 $\pm\ $0.0022& 5.0 & 50 & 3.42 & 8.63 & 9.09 & --18.30 \\
\hline
Kes 69 Maser
& +69.76 $\pm\ $0.22& 1.91 $\pm\ $0.12& 16.5 $\pm\ $0.3& 0.0013 $\pm\ $0.0001 & 5.0 & 75 &  1.32 & 3.70 & 3.36 & --6.22\\
Kes 69 Shell
& +85.2  $\pm\ $0.6 & 1.3  $\pm\ $0.7 & 16.49 $\pm\ $1.06 & 0.005 $\pm\ $0.001& 5.0 & 40 & 3.44 & 13.98 & 11.99 & --8.21 \\ 
& +83.6  $\pm\ $0.9 & 6.1 $\pm\ $1.5 & 16.23 $\pm\ $0.65  & 0.091 $\pm\ $0.013& 5.0 & 30 & 6.47 & 21.52 & 18.01 & --21.67 \\
\hline
3C391
 & +109.62 $\pm\ $0.15& 1.40 $\pm\ $0.15& 17.1 $\pm\ $0.4& 0.0009 $\pm\ $0.0001& 5.0& 65 & 2.60 & 6.30 & 5.98 & --22.18 \\ 
 & +108.52 $\pm\ $0.11& 2.34 $\pm\ $0.09& 16.3 $\pm\ $0.1& 0.022 $\pm\ $0.001  & 5.0& 30& 16.64 & 17.34 & 17.74 & 18.48\\
 & +105.30 $\pm\ $0.15& 1.50 $\pm\ $0.16& 16.4 $\pm\ $0.1& 0.0041 $\pm\ $0.0004& 5.0& 40& 3.19 & 15.07 & 11.90 & --6.76\\
 & +97.04 $\pm\ $0.05 & 3.69 $\pm\ $0.03& 16.0 $\pm\ $0.1& 0.043 $\pm\ $0.001  & 5.0& 50 & 1.82 & 10.44 & 7.28 & --3.46\\
\hline
W44 A
& +42.95 $\pm\ $0.01& 2.31 $\pm\ $0.01& 17.13 $\pm\ $0.01& 0.010 $\pm\ $0.001& 5.0 & 75 & 2.37 & 5.40 & 4.70 & --7.76\\
& +44.6  $\pm\ $0.1 & 2.18 $\pm\ $0.03& 16.6  $\pm\ $0.4 & 0.003 $\pm\ $0.001& 5.0 & 100& 1.01 & 2.69 & 2.33 & --6.70\\
& +39.3  $\pm\ $0.2 & 3.2  $\pm\ $0.2 & 16.09 $\pm\ $0.05& 0.012 $\pm\ $0.001& 5.0 & 50 & 6.79 & 7.28 & 7.54 & 8.10\\
& +32.2  $\pm\ $0.3 & 12.4 $\pm\ $0.2 & 16.54 $\pm\ $0.03& 0.017 $\pm\ $0.001& 5.0 & 60 & 4.67 & 4.92 & 5.02 & 5.29\\
W44 B,C
& +45.68 $\pm\ $0.01& 1.98 $\pm\ $0.11& 17.13 $\pm\ $0.01& 0.0020 $\pm\ $0.0001& 5.0& 75& 2.37 & 5.40 & 4.70 & --7.31\\
& +44.80 $\pm\ $0.01& 1.46 $\pm\ $0.01& 17.13 $\pm\ $0.01& 0.0010 $\pm\ $0.0001& 5.0& 75& 2.37 & 5.40 & 4.70 & --7.52\\
& +43.33 $\pm\ $0.06& 6.00 $\pm\ $0.05& 16.82 $\pm\ $0.03& 0.0107 $\pm\ $0.0001& 5.0& 75& 1.58 & 4.05 & 3.79 & --10.18\\
& +39.50 $\pm\ $0.12& 11.2 $\pm\ $0.1&  15.9 $\pm\ $0.1 &  0.0646 $\pm\ $0.0003& 5.0& 40& 10.42& 11.08& 11.47& 12.21\\
& +38.90 $\pm\ $0.09& 2.97 $\pm\ $0.07& 16.4 $\pm\ $0.2 &  0.0180 $\pm\ $0.0002& 5.0& 40& 10.51& 11.18& 11.57& 12.32\\
& +23.8 $\pm\ $0.1  & 37.6 $\pm\ $0.2 & 16.6 $\pm\ $0.1 &  0.066 $\pm\ $0.001&   5.0& 30& 16.69& 17.42& 17.82& 18.59\\
W44 D,E,F
& +40.06 $\pm\ $0.22  & 5.58 $\pm\ $0.10  & 15.95 $\pm\ $0.03& 0.137 $\pm\ $0.011  & 5.0& 50& 1.81 & 10.45& 7.26 & --3.43\\
& +44.61 $\pm\ $0.02& 2.47 $\pm\ $0.03& 17.14 $\pm\ $0.01& 0.0101 $\pm\ $0.0003& 5.0& 75& 2.38 & 5.40 & 4.71 & --6.42\\
& +47.35 $\pm\ $0.02& 1.60 $\pm\ $0.11& 17.13 $\pm\ $0.01& 0.0032 $\pm\ $0.0002& 5.0& 75& 2.36 & 5.40 & 4.70 & --8.22\\
& +45.8 $\pm\ $0.2  & 4.7 $\pm\ $0.2  & 16.00 $\pm\ $0.03& 0.102 $\pm\ $0.002  & 5.0& 50& 6.77 & 7.26 & 7.53 & 8.10\\
& +37.7 $\pm\ $0.2  & 25.6 $\pm\ $0.4 & 16.38 $\pm\ $0.05& 0.094 $\pm\ $0.001  & 5.0& 60& 4.61 & 4.88 & 4.99 & 5.28\\
\hline
W51C 
 & +72.13 $\pm\ $0.06& 1.12 $\pm\ $0.08& 16.49 $\pm\ $0.31& 0.0045 $\pm\ $0.0006& 5.0 & 75 & 1.22 & 3.66 & 3.23 & --4.90\\
 & +69.19 $\pm\ $0.03& 1.17 $\pm\ $0.04& 16.89 $\pm\ $0.20& 0.0071 $\pm\ $0.0006& 5.0 & 75 & 1.70 & 4.28 & 4.01 & --12.19\\
 & +65.70 $\pm\ $0.04& 4.54 $\pm\ $0.09& 15.41 $\pm\ $0.02& 0.59 $\pm\ $0.01& 5.0 & 40 & 10.47 & 11.06 & 11.42 & 12.07\\
\hline
IC443 B
&  --6.1 $\pm\ $0.1&  1.8 $\pm\ $0.1& 15.95 $\pm\ $0.02& 0.018 $\pm\ $0.001& 5.0 & 50 & 1.80 & 10.43 & 7.19 & --3.40 \\
& --14.3 $\pm\ $0.2& 17.2 $\pm\ $0.6& 16.05 $\pm\ $0.02& 0.15 $\pm\ $0.01&   5.0 & 50 & 6.78 & 7.27 & 7.54 & 8.10\\
& --32.0 $\pm\ $1.5& 37.9 $\pm\ $1.6& 16.50 $\pm\ $0.03& 0.37 $\pm\ $0.28&   5.0 & 50 & 6.95 & 7.39 & 7.63 & 8.11\\
IC443 D
& --6.80 $\pm\ $0.03& 2.30 $\pm\ $0.11& 16.38 $\pm\ $0.04& 0.008 $\pm\ $0.004& 5.0 & 50 & 2.14 & 8.59 & 7.32 & --5.23\\
& --12.7 $\pm\ $1.4&  21.2 $\pm\ $3.3&  16.05 $\pm\ $0.02& 0.560 $\pm\ $0.075&   5.0 & 40 & 10.43 & 11.11 & 11.49 & 12.25\\
IC443 G
& --4.60 $\pm\ $0.02& 1.38 $\pm\ $0.13& 17.11 $\pm\ $0.03& 0.015 $\pm\ $0.006 & 5.0 & 75 & 2.27 & 5.37 & 4.69 & --16.62\\
& --5.47 $\pm\ $0.95& 17.0 $\pm\ $2.3&  16.1 $\pm\ $0.1&   0.60 $\pm\ $0.07   & 5.0 & 40 & 10.43 & 11.11 & 11.50 & 12.25\\
\hline
CTB 37A 
& --66.2 $\pm\ $0.1& 1.3 $\pm\ $0.1  & 15.86 $\pm\ $0.03& 0.0052 $\pm\ $0.0005& 5.0 & 100 & 0.68 & 2.49 & 1.86 & --2.17\\
& --63.5 $\pm\ $0.1& 1.53 $\pm\ $0.08& 16.36 $\pm\ $0.02& 0.0049 $\pm\ $0.0003& 5.0 & 100 & 0.86 & 2.57 & 2.11 & --3.90\\
G348.5--0.0
& --21.3 $\pm\ $0.1& 1.1 $\pm\ $0.1  & 17.02 $\pm\ $0.03& 0.0009 $\pm\ $0.0004& 5.0 & 75  & 1.96 & 4.84 & 4.41 & --17.47\\
& --22.0 $\pm\ $0.2& 4.5 $\pm\ $0.4  & 15.51 $\pm\ $0.02& 0.018 $\pm\ $0.001&   5.0 & 75  & 2.85 & 3.03 & 3.08 & 3.28\\
\hline
G349.7+0.2
 & +16.80 $\pm\ $0.04 & 1.16 $\pm\ $0.08 & 16.99 $\pm\ $0.04 & 0.005 $\pm\ $0.001 & 5.0 & 75 & 1.90 & 4.70 & 4.33 & --15.67 \\
 & +15.48 $\pm\ $0.02 & 1.19 $\pm\ $0.11 & 17.04  $\pm\ $0.02 & 0.007 $\pm\ $0.001  & 5.0 & 75 & 1.92 & 4.74 & 4.36 & --16.04 \\
\hline
Square
& --35.40 $\pm\ $0.03& 6.98 $\pm\ $0.02& 17.12 $\pm\ $0.01& 0.091 $\pm\ $0.001& 5.0 & 75 & 2.32 & 5.38 & 4.70 & --12.63\\
& --35.63 $\pm\ $0.02& 2.72 $\pm\ $0.03& 16.68 $\pm\ $0.07& 0.053 $\pm\ $0.001& 5.0 & 65 & 1.71 & 4.66 & 4.47 & --8.08\\
& --38.55 $\pm\ $0.06& 33.3 $\pm\ $0.1 & 15.56 $\pm\ $0.02& 0.574 $\pm\ $0.003  & 5.0 & 75 & 2.85 & 3.03 & 3.09 & 3.28\\
\hline
Tornado
& --12.7 $\pm\ $0.1& 1.7 $\pm\ $0.1& 15.90 $\pm\ $0.09& 0.014 $\pm\ $0.002& 5.0 & 125 & 0.60 & 1.72 & 1.38 & --3.48\\
& --11.8 $\pm\ $0.1& 1.5 $\pm\ $0.1& 15.89 $\pm\ $0.11& 0.015 $\pm\ $0.001& 5.0 & 125 & 0.60 & 1.72 & 1.38 & --3.46\\
& --13.8 $\pm\ $0.1& 1.5 $\pm\ $0.1& 15.89 $\pm\ $0.12& 0.020 $\pm\ $0.002& 5.0 & 125 & 0.60 & 1.72 & 1.38 & --3.46\\
& --15.1 $\pm\ $0.1& 1.6 $\pm\ $0.1& 15.89 $\pm\ $0.12& 0.016 $\pm\ $0.002& 5.0 & 125 & 0.60 & 1.72 & 1.38 & --3.46\\
& --13.2 $\pm\ $0.2& 6.5 $\pm\ $0.3& 15.29 $\pm\ $0.05& 0.020 $\pm\ $0.003& 5.0 & 125 & 1.30 & 1.36 & 1.34 & 1.40 \\
\hline
G359.1--0.5
& --5.24 $\pm\ $0.02& 1.74 $\pm\ $0.09& 17.13 $\pm\ $0.01& 0.0005 $\pm\ $0.0001& 5.0 & 75 & 2.37 & 5.40 & 4.71 & --7.10\\
& --6.04 $\pm\ $0.06& 45.3 $\pm\ $0.1& 17.11 $\pm\ $0.01& 0.102 $\pm\ $0.003& 5.0 & 20 & 18.58 & 18.76 & 18.84 & 19.01\\
& --11.36 $\pm\ $0.08& 4.69 $\pm\ $0.05& 16.85 $\pm\ $0.03& 0.030 $\pm\ $0.002& 5.0 & 20 & 18.57 & 18.73 & 18.80 & 18.95\\
& --12.52 $\pm\ $0.04& 31.5 $\pm\ $0.1& 16.02 $\pm\ $0.03& 0.095 $\pm\ $0.001& 5.0 & 100 & 0.73 & 2.53 & 1.92 & --2.48\\
& --23.68 $\pm\ $0.07& 51.0 $\pm\ $0.1& 16.94 $\pm\ $0.03& 0.101 $\pm\ $0.002& 5.0 & 20 & 7.30 & 7.65 & 7.82 & 8.19 \\
\hline
\end{tabular}
\tablecomments{We include the excitation temperature for each transition for reference, but this is not a free parameter\\ and is directly determined by the models.}
\end{table}

\begin{table}
\centering
\caption{Anomalous 1720 Clouds Best-Fit Parameters \label{tbl:anomfits}}
\begin{tabular}{lrrrrccccccccccccccccc}
\hline \hline
{SNR} &
{$v_0$} & 
{$\Delta v$} & 
{log $\noh$} & 
{$f$} & 
{log $n$} & 
{T$_k$} & 
{$\tex^{1612}$} &  
{$\tex^{1665}$} &  
{$\tex^{1667}$} & 
{$\tex^{1720}$} \\
{Name} & {(\kms )} & {(\kms )} &{ $({\rm cm}^{-2})$} & & {(cm$^{-3}$)} & {(K)} & {(K)} & {(K)} & {(K)} & {(K)}\\
\hline
49.2-0.7 & +61.00 $\pm\ $0.06& 3.01 $\pm\ $0.13& 15.47 $\pm\ $0.06& 0.127 $\pm\ $0.005& 3.0 & 50 & 1.65 & 9.60 & 6.37 & --3.18\\
348.5-0.0 & --9.47 $\pm\ $0.04& 3.15 $\pm\ $0.09& 15.77 $\pm\ $0.02& 0.07 $\pm\ $0.01&    2.0 & 40  & 2.81 & 10.73 & 8.67 & --7.60\\
349.7+0.2 & --60.7 $\pm\ $0.6  & 5.8  $\pm\ $0.8  & 17.5  $\pm\ $0.1 & 0.011 $\pm\ $0.002 & 2.5 & 25 & 9.57 & 20.95 & 18.51 & 1521.23 \\
 & --63.7 $\pm\ $0.2  & 3.6  $\pm\ $0.3  & 15.6  $\pm\ $0.7 & 0.12  $\pm\ $0.11   & 1.5 & 25 & 7.89 & 11.57 & 12.36 & 23.08 \\
 & --95.4 $\pm\ $0.3  & 6.6  $\pm\ $0.5  & 17.55 $\pm\ $0.06 & 0.014 $\pm\ $0.001& 2.5 & 25 & 9.58 & 20.97 & 18.54 & 1630.19 \\
 &--110.0 $\pm\ $0.4  & 4.9  $\pm\ $0.5  & 14.99 $\pm\ $0.07 & 1.0 $\pm\ $0.2    & 2.5 & 25 & 9.20 & 17.64 & 16.41 & 82.82 \\
\hline
\end{tabular}
\tablecomments{We include the excitation temperature for each transition for reference, but this is not a free parameter\\ and is directly determined by the models.}
\end{table}

\begin{table} \centering
\caption{Integrated Line Intensity of OH(1720 MHz) Emission \label{tbl:eme}}
\begin{tabular}{rrccccr}
\hline \hline
{SNR} & {Name} & {Velocity Range} & {GBT 0$^{th}$ Moment} & {VLA 0$^{th}$ Moment} & {GBT/VLA} & {Ref.} \\
 & & {(\kms)} & {(Jy \kms)}  & {(Jy \kms)} & {Ratio}\\
\hline
6.4--0.1  & W28 A,B  & +4.8,+6.3 & 15.3 & 7.83 & 2.0 & C97 \\ 
          & W28 C,D & +7.2,+14.1 & 11.6 & 6.98 & 1.7 & C97 \\ 
          & W28 E,F & +8.6,+15.9 & 273 & 176 & 1.6 & C97 \\ 
16.7+0.1  &         & +19.9 & 0.41 & 0.25 & 1.7 & G97\\
21.8--0.6 & Kes 69  & +69.8 & 0.13 & 0.093 & 1.4 & G97\\
 31.9+0.0 & 3C391 & +105.3 & 0.20 & 0.063 & 3.2 & F96\\
         &      & +109.8 & 0.26 & 0.077 & 3.3 & F96\\
34.7--0.4 & W44 A  & +42.9       & 7.64 & 1.29 & 5.9 & C97 \\ 
          & W44 B,C & +44.7,+46.7 & 17.9 & 2.31 & 7.7 & C97 \\ 
          & W44 D,E,F  & +43.7,+46.9 & 48.9 & 33.4 & 1.5 & C97 \\ 
49.2--0.7 & W51C    & +67.4,+74.3 & 7.82 & 8.31 & 0.9 & B00 \\
189.1+3.0 & IC 443 G& --4.6,0.5 & 3.92 & 3.17 & 1.2 & C97\\
          & IC 443 D& --9.9,--4.9 & 0.29 & -- & -- & \\ %new 
          & IC 443 B& --12.6,--4.9 & 0.22 & -- & -- & \\ %new 
348.5--0.0 &         & --21.4 & 0.33 & 0.38 & 0.9 & B00 \\
348.5+0.1 & CTB 37A & --63.5,--66.2 & 1.67 & 1.70 & 1.0 & B00 \\
349.7+0.2 &         & +14.3,+16.7 & 3.53 & 2.15 & 1.6 & B00 \\
357.7--0.1 & Tornado & --12.8,--14.7 & 3.01 & 0.59 & 5.1 & F99 \\
357.7+0.3 & Square  & --34.1,--37.2 & 9.43 & 1.12 & 8.4 & F99 \\
\hline
\end{tabular}
\tablecomments{References: F96=\cite{frail96}; C97=\cite{claussen97}; G97=\cite{green97}; K98=\cite{koralesky98}; F99=\cite{fyz99}; B00=\cite{brogan00}.}
\end{table}

\clearpage

\clearpage

\begin{figure} \centering
\includegraphics[angle=0,width=5.in]{f9}
\caption{[Left] 20cm Radio continuum image of SNR W44. Contours are drawn at levels of 78, 98 and 120 mJy \beam (synthesized beam of 15\arcsec ). [Right] Integrated line intensity of 1720 MHz maser emission towards the W44 in units of Jy \kms . Maser groups A through F are labeled along contours of the radio shell.
\label{fig:mom-w44}}
\end{figure}

\begin{figure}
\centering
\includegraphics[angle=-90,width=3.9in]{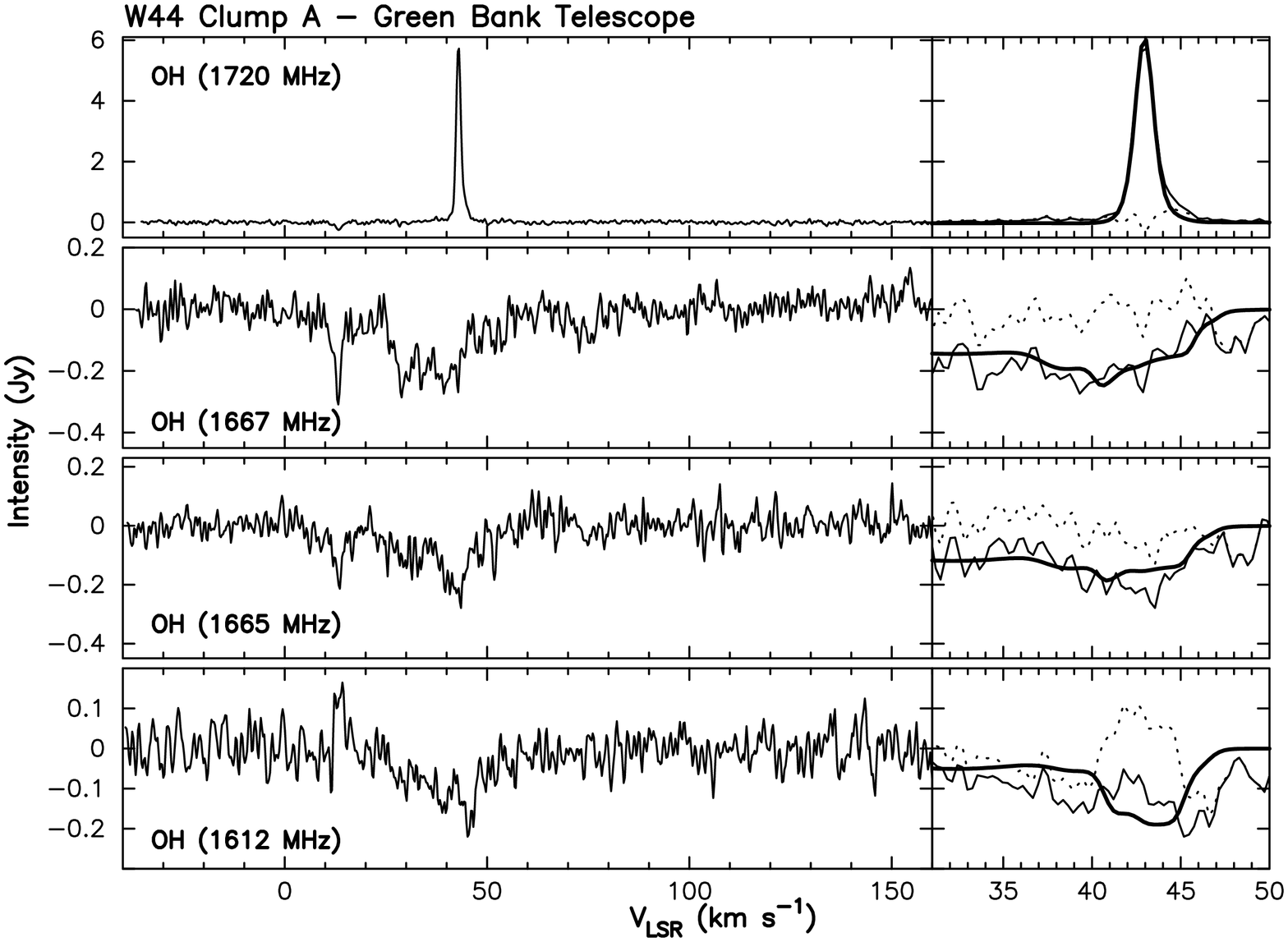}\bigskip \\
\includegraphics[angle=-90,width=3.9in]{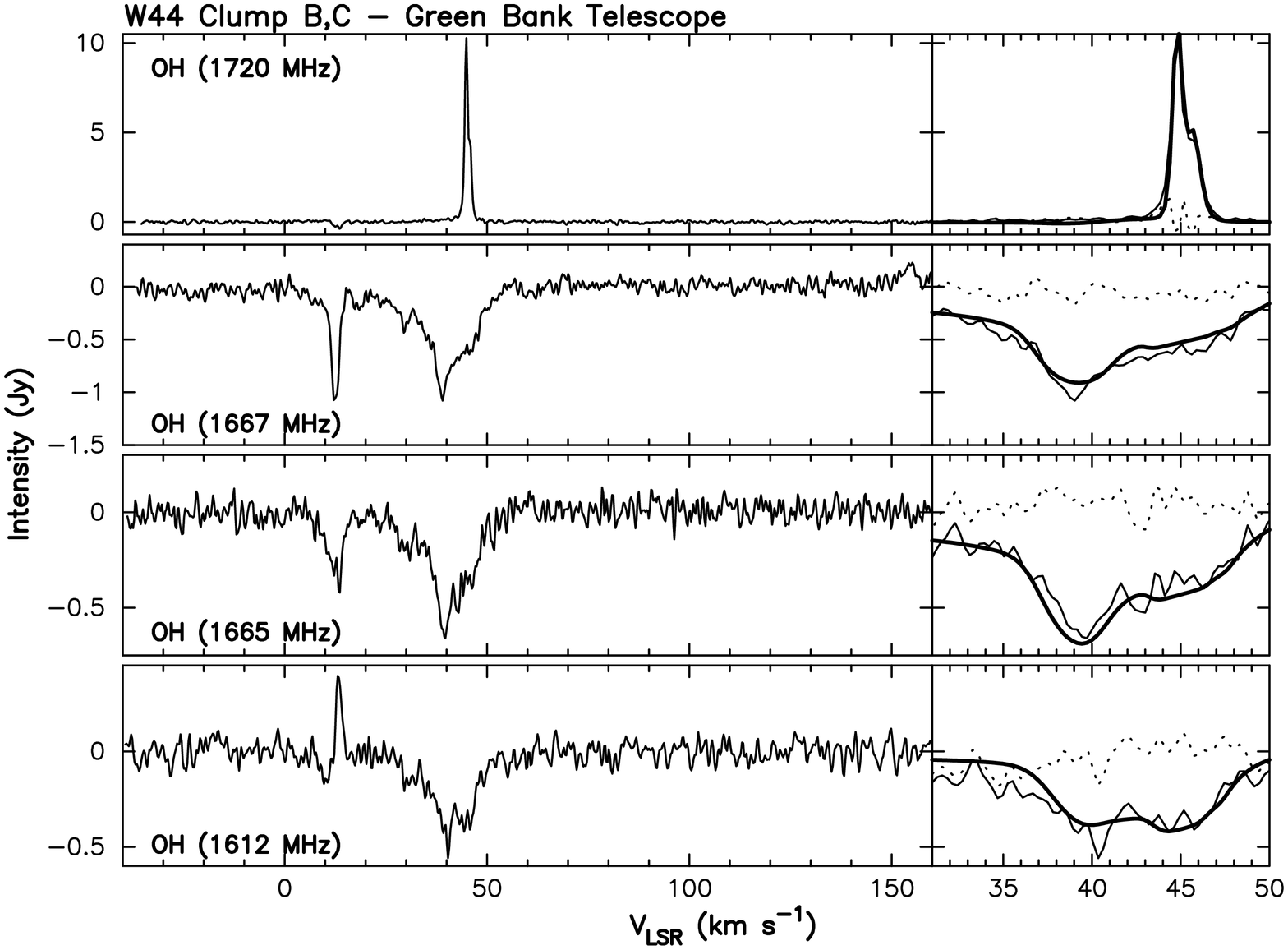}\bigskip \\
\includegraphics[angle=-90,width=3.9in]{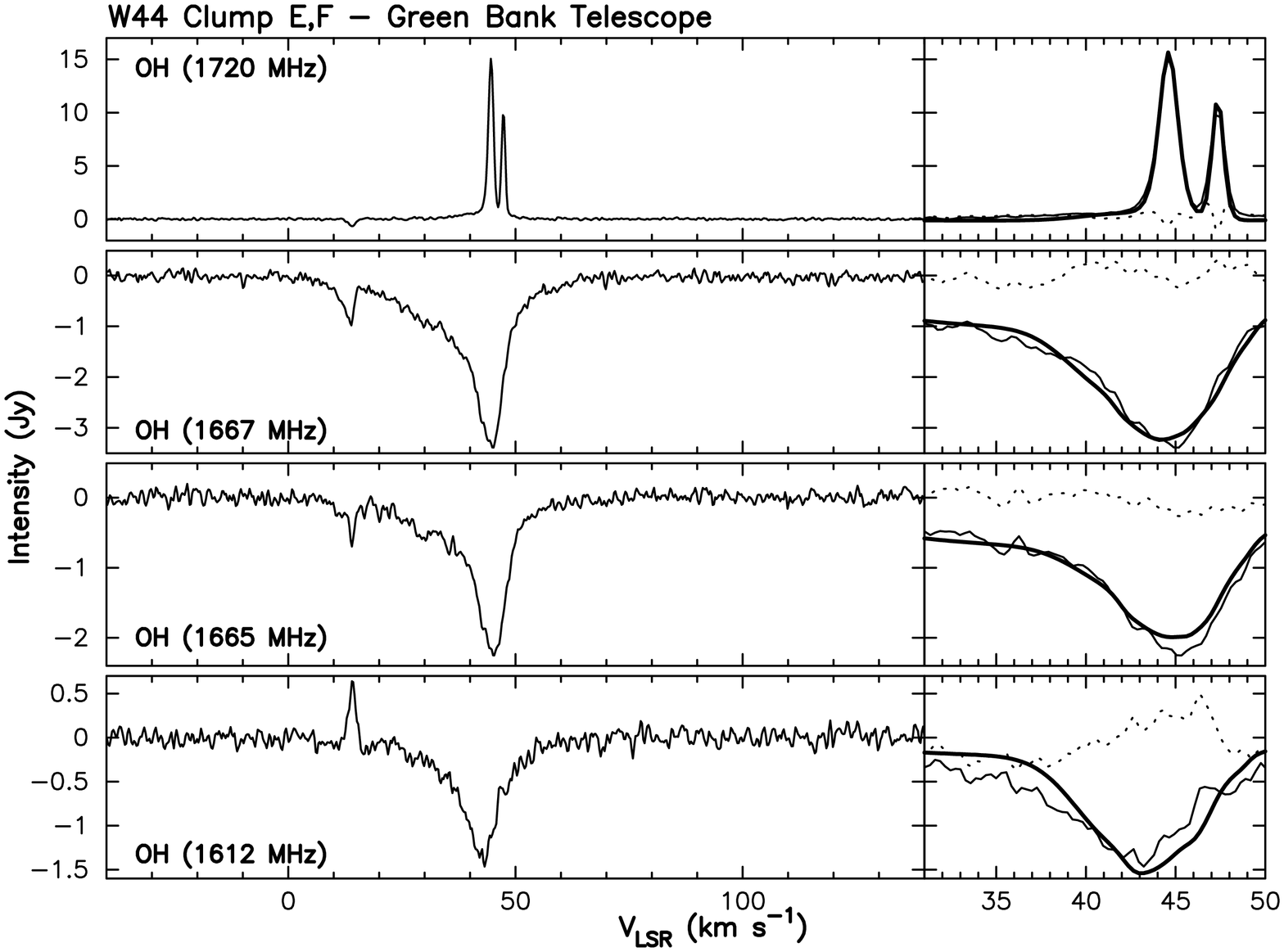}
\caption{Spectra of all four OH transitions towards W44 as in Figure \ref{fig:spec-w28}. Clump A shows emission at +43\kms , with accompanying broad absorption. Clumps B and C show emission at +44.7 and +45.7 \kms , with accompanying broad absorption centered at +40\kms . Clumps E and F shows two narrow maser components and a broad emission feature. \label{fig:spec-w44}}
\end{figure}

\begin{figure} \centering
\includegraphics[angle=-90,width=3.5in]{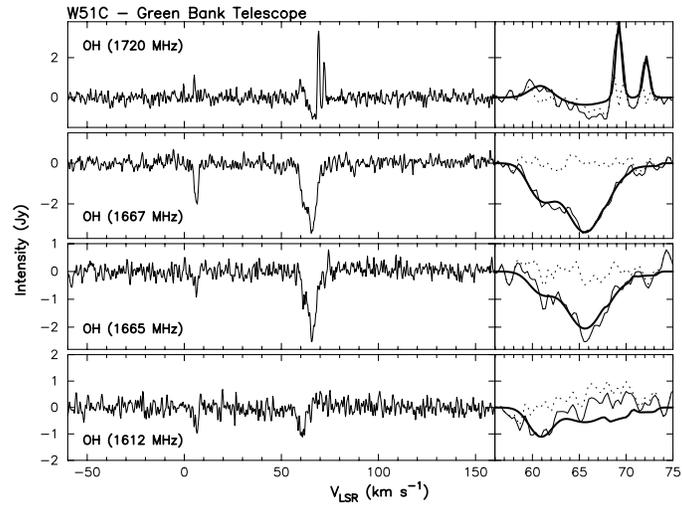}
\caption{Spectra of all four OH transitions towards W51C  as in Figure \ref{fig:spec-w28} centered at $\alpha$,$\delta$(J2000) = 19 22 57.3, +14 15 52. \label{fig:spec-w51c}}
\end{figure}

\begin{figure} \centering
\includegraphics[angle=90,width=3.5in]{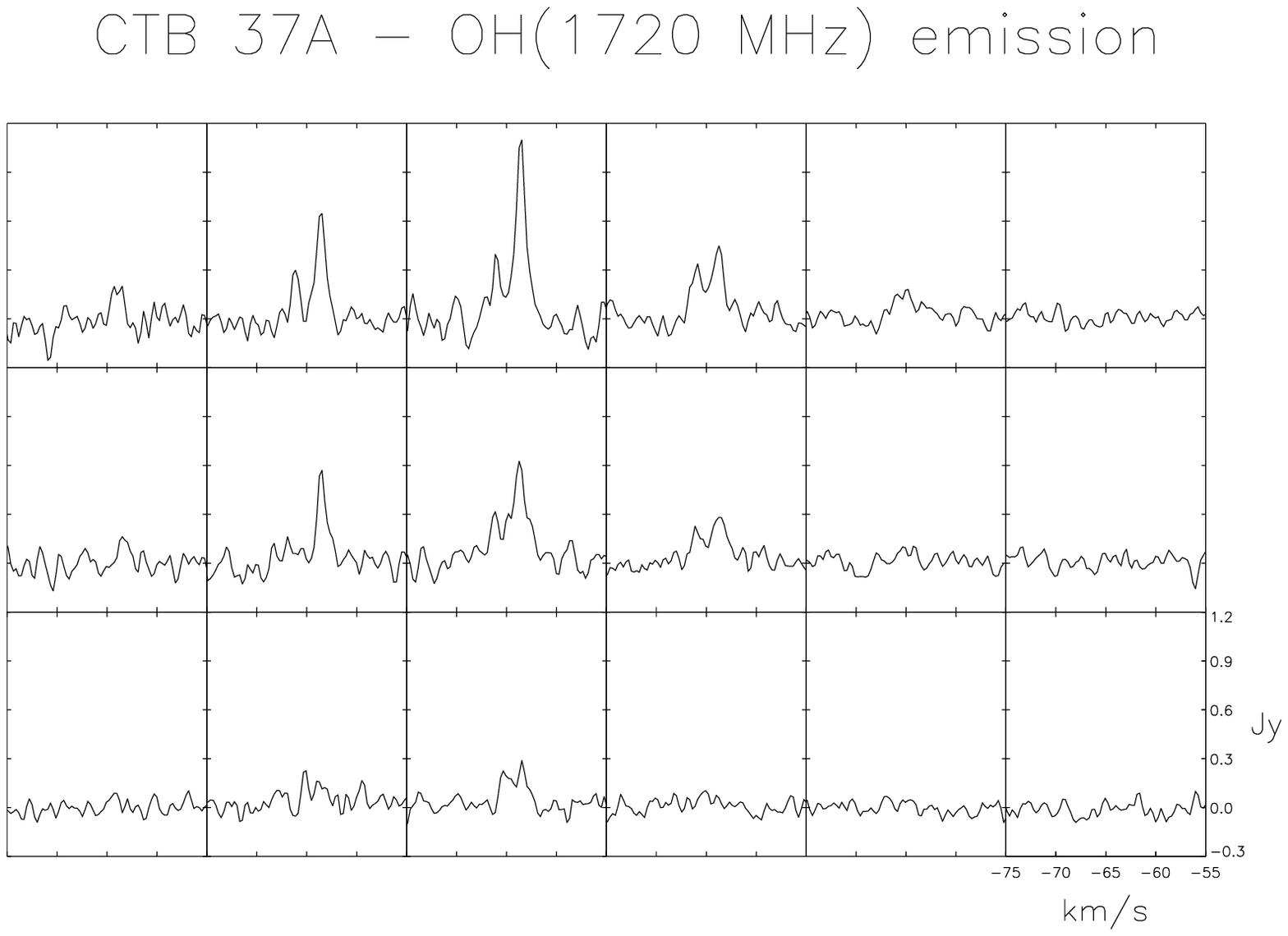}
\caption{Postage stamp plot of OH(1720 MHz) maser emission between -55 and -75\kms\ from CTB 37A. Map is centered at $\alpha$,$\delta$(J2000) = 6$^{\rm h}$16$^{\rm m}$45$\dsec$4, +22$\degr$32$\arcmin$16$\arcsec$ with pointings spaced by 3$\damin$3 in the RA and DEC directions. 
Two masers are seen at --66.2 and --63.5\kms .\label{fig:post-ctb37a}}
\end{figure}

\begin{figure} \centering
\includegraphics[angle=90,width=3.8in]{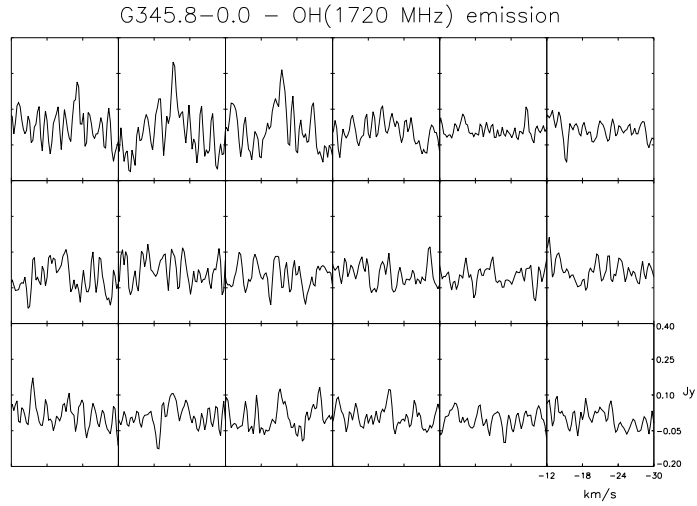}
\caption{
Postage stamp plot of OH(1720 MHz) maser emission between -12 and -30\kms\ from G348.5-0.0. Map is centered at $\alpha$,$\delta$(J2000) = 6$^{\rm h}$16$^{\rm m}$45$\dsec$4, +22$\degr$32$\arcmin$16$\arcsec$ with pointings spaced by 3$\damin$3 in the RA and DEC directions. A maser is seen in the north eastern corner at --21.3\kms \label{fig:post-g3485}}
\end{figure}

\begin{figure} \centering
\includegraphics[angle=-90,width=3.5in]{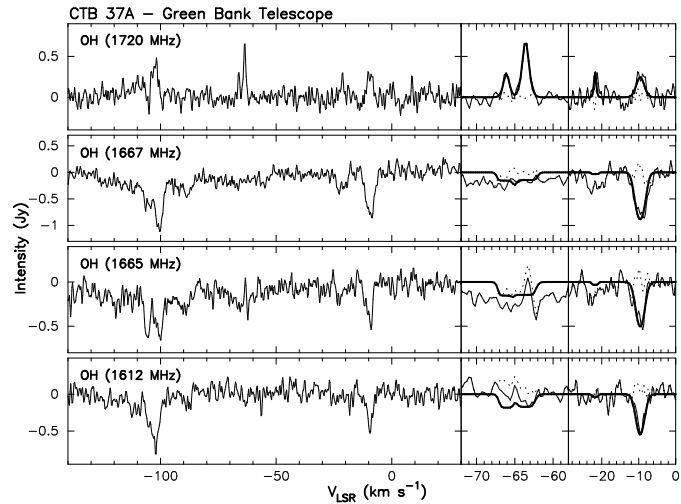}
\caption{Spectra of all four OH transitions towards CTB 37A masers as in Figure \ref{fig:spec-w28}. 
Two masers are seen at --66.2 and --63.5\kms associated with CTB 37A, and one maser is seen at --21.3\kms\ associated with G348.5-0.0. A giant molecular cloud is seen at -10\kms . \label{fig:spec-ctb37a}}
\end{figure}

\begin{figure} \centering
\includegraphics[angle=-90,width=3.5in]{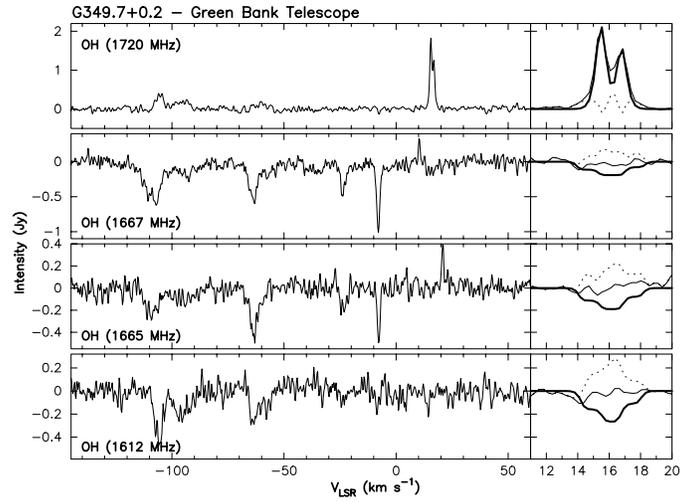}
\caption{Spectra towards G349.7+0.2 centered at 17h 14m 48.0s, -37d 20m 39s (J2000) as in Figure \ref{fig:spec-w28}.
Emission at +15.5 and +16.8 \kms\ shows a significantly increased flux in comparison to high resolution VLA observations, though no corresponding absorption is seen at these velocities.
\label{fig:spec-g3497}}
\end{figure}

\begin{figure} \centering
\includegraphics[angle=-90,width=3.5in]{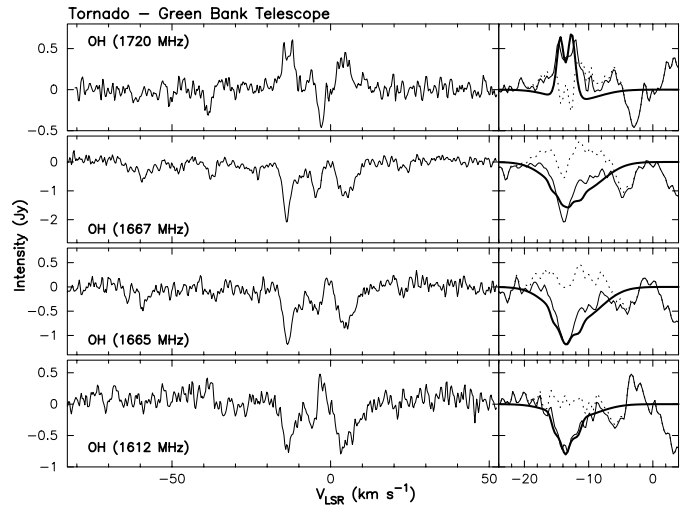}
\caption{Spectra towards the Tornado SNR at 17h 40m 02.8s , -30d 58m 51.4s (J2000) as in Figure \ref{fig:spec-w28}.
 \label{fig:spec-tornado}}
\end{figure}

\begin{figure} \centering
\includegraphics[angle=0,width=4.6in]{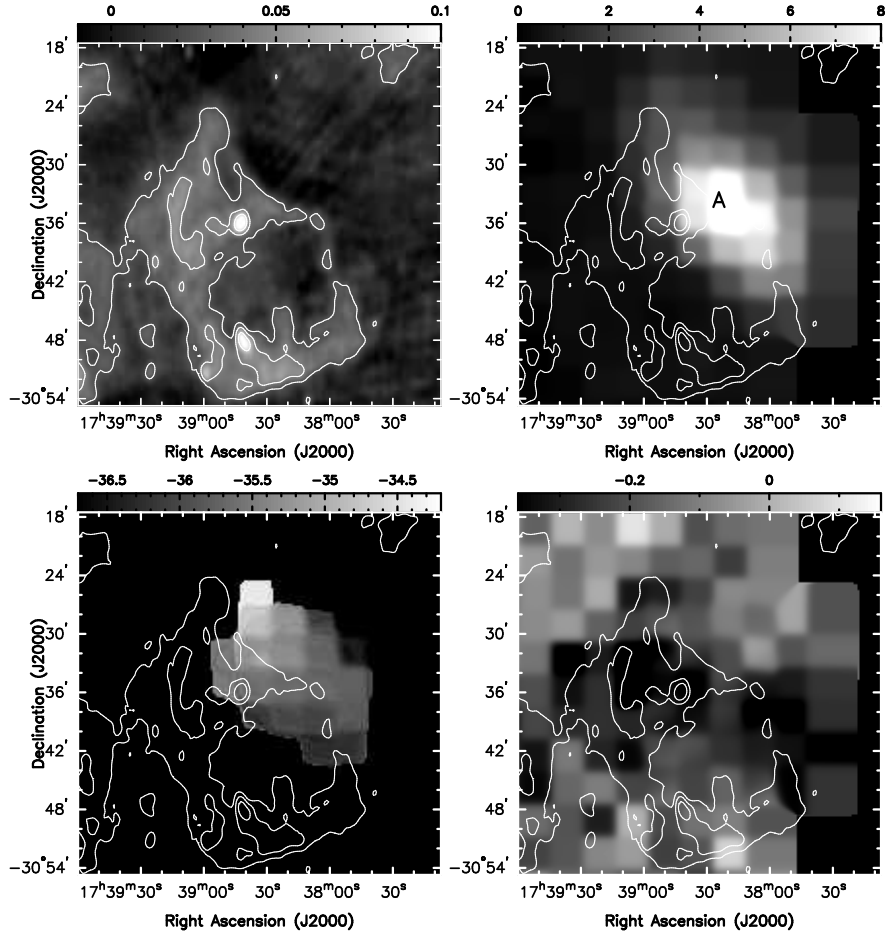}
\caption{[Top Left] 20cm Radio continuum image of the Square Nebula (G357.7+0.3). Contours are drawn at levels of 25, 45 and 80 mJy \beam (synthesized beam of 15\arcsec ). [Top Right] Integrated line intensity of 1720 MHz maser emission shows emission extending across the western extent of the SNR in units of Jy \kms . [Bottom Left] A map of the first moment of maser-line emission shows a velocity gradient from -36.8 to -34.2\kms\ across the extended maser structure. [Bottom Right] A foreground cloud is revealed across the bright continuum of the remnant in this integrated line intensity of 1720 MHz absorption centered at 0 \kms . Units are Jy \kms .\label{fig:mom-square}}
\end{figure}

\begin{figure} \centering
\includegraphics[angle=-90,width=4.1in]{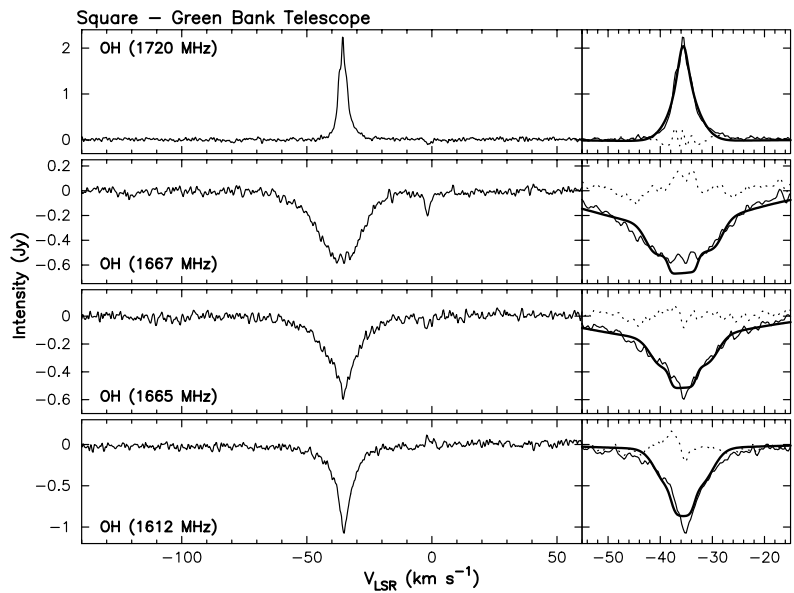}
\caption{Spectra towards the Square nebula SNR at 17h 38m 20.1s , -30d 33m 58.9s
(J2000) as in Figure \ref{fig:spec-w28}. \label{fig:spec-square}}
\end{figure}

\begin{figure} \centering
\includegraphics[angle=0,width=4.5in]{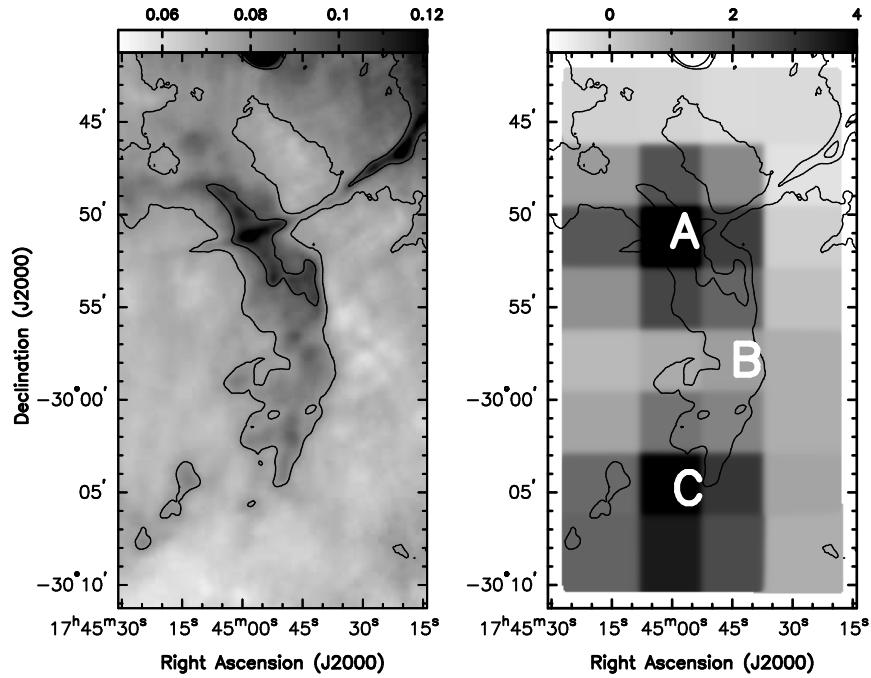}
\caption{[Left] 20cm Radio continuum image of SNR G359.1-0.5. Contours are drawn at levels of 78, 98 and 120 mJy \beam (synthesized beam of 30\arcsec ). [Right] Integrated line intensity of 1720 MHz maser emission towards the Square Nebula in units of Jy \kms . Maser groups A, B and C are labeled along contours of the radio shell.
\label{fig:mom-g3591}}
\end{figure}

\begin{figure} \centering
\includegraphics[angle=-90,width=4.5in]{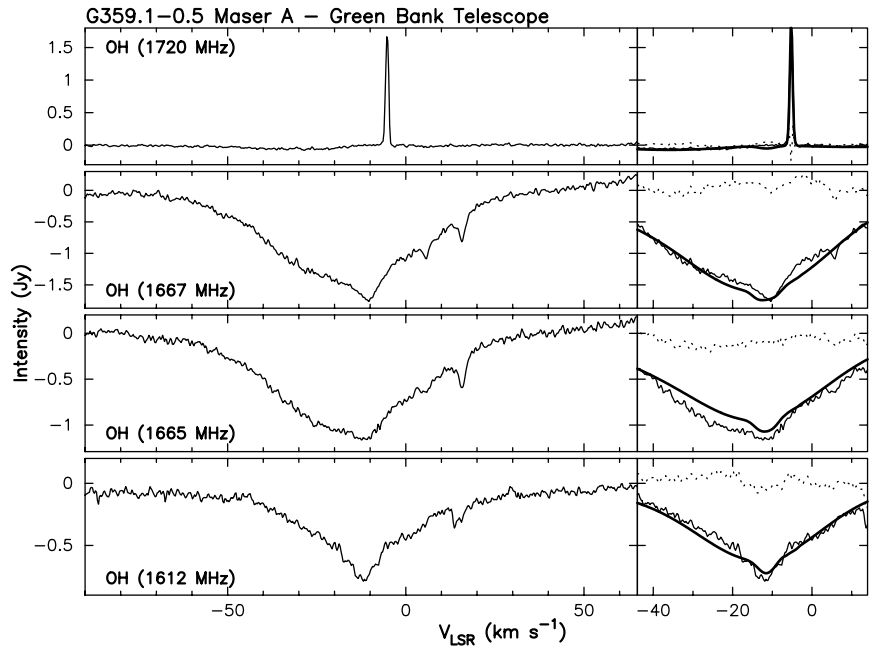}
\caption{Spectra towards Maser A of SNR G359.1-0.5 as in Figure \ref{fig:spec-w28}.
 \label{fig:spec-g3591a}}
\end{figure}

\begin{figure} \centering
\includegraphics[angle=-90,width=6in]{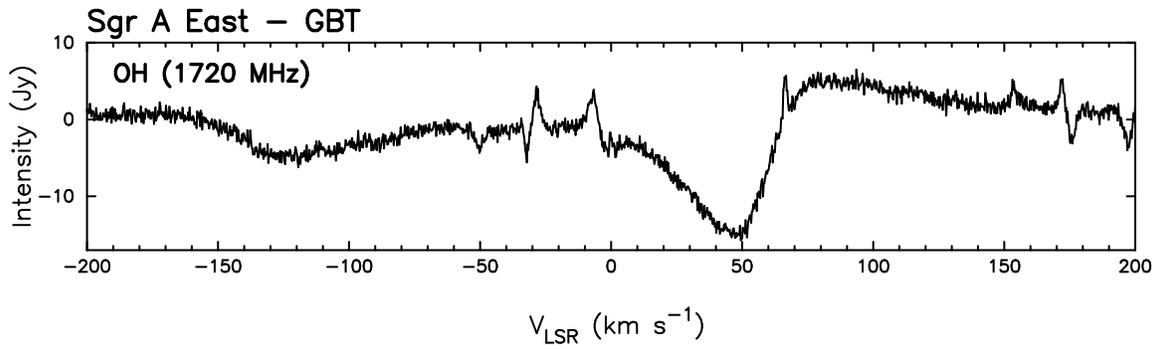}
\caption{The 1720 MHz spectrum of Sgr A East towards the brightest 10 Jy maser. Several sources of confusion are present, particularly wide, deep absorption that makes analyzing maser emission from the SNR impractical. Similar confusion is present toward SNRs G1.0-0.1 and G1.4-0.1. \label{fig:spec-sgra}}
\end{figure}

\begin{figure} \centering
\includegraphics[angle=90,width=7in]{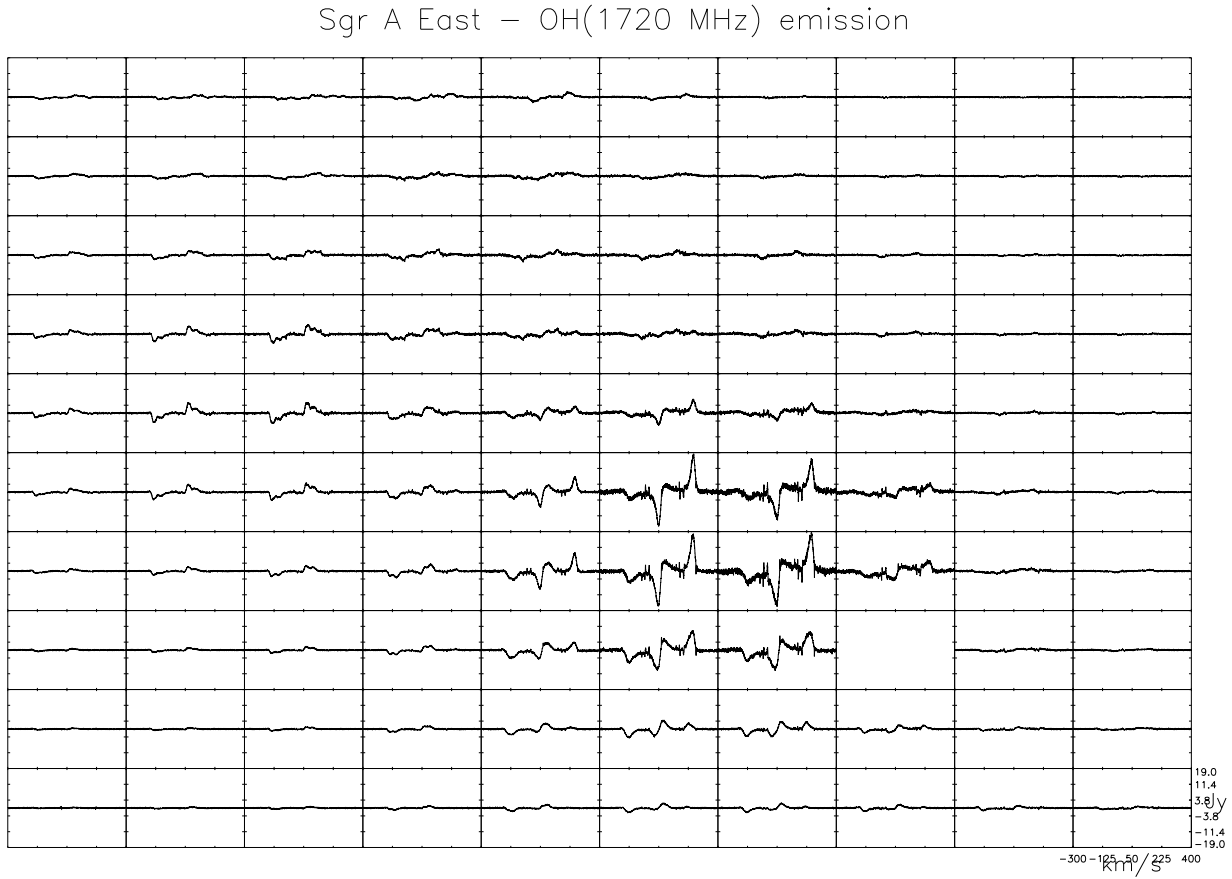}
\caption{Postage stamp plot of OH(1720 MHz) maser emission and absorption towards Sgr A East. Spectra were taken in a 10 by 10 grid spaced by 3$\damin$3 in the RA and DEC directions. Spectra are frequency switched so the strong absorption seen at +50 to --150\kms\ appears in conjugate at +100 to +300\kms . One pointing was corrupted by interference and is left blank. \label{fig:post-sgra}}
\end{figure}

\begin{figure} \centering
\includegraphics[scale=0.41,angle=90]{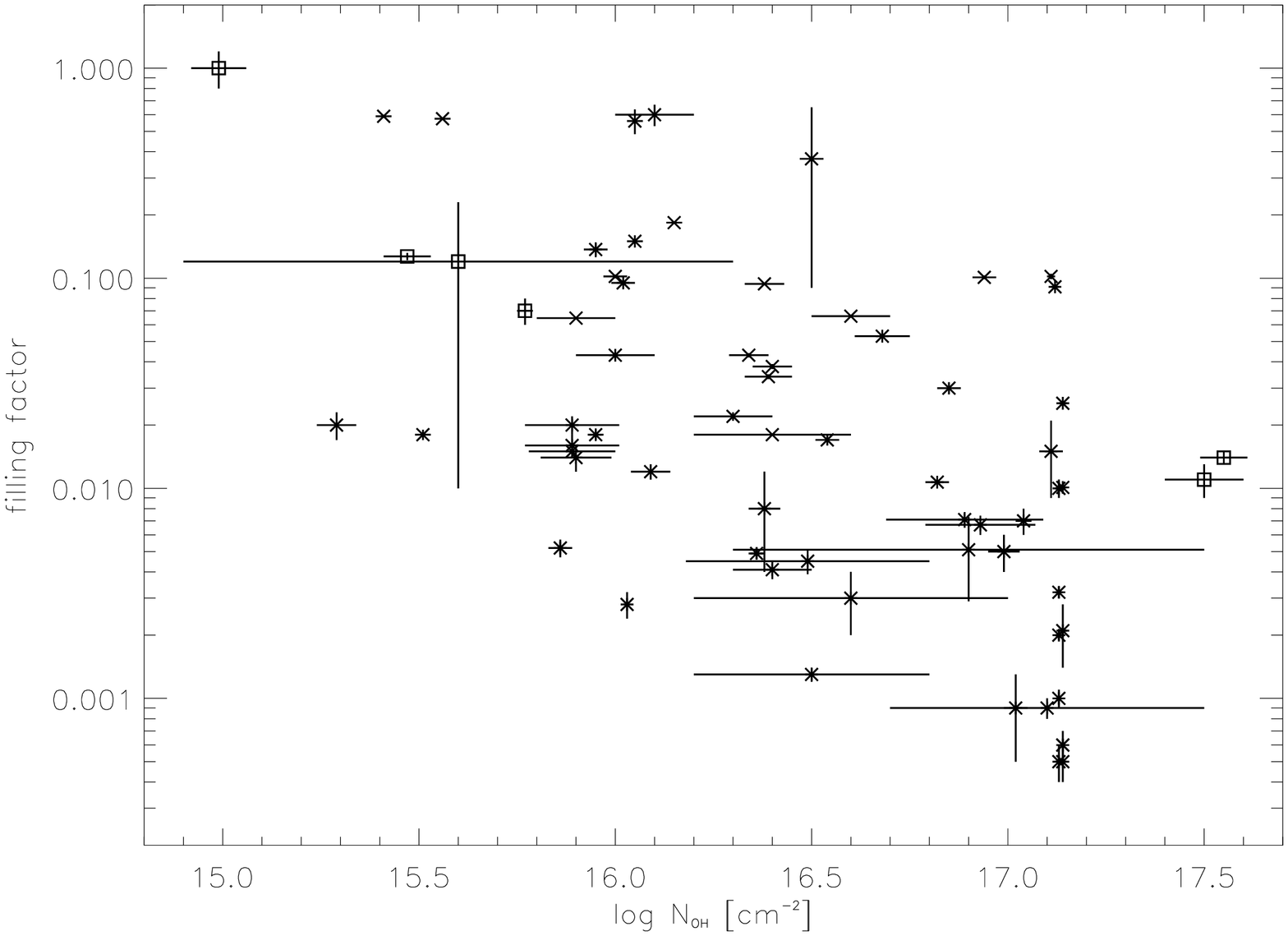}\\
\includegraphics[scale=0.41,angle=90]{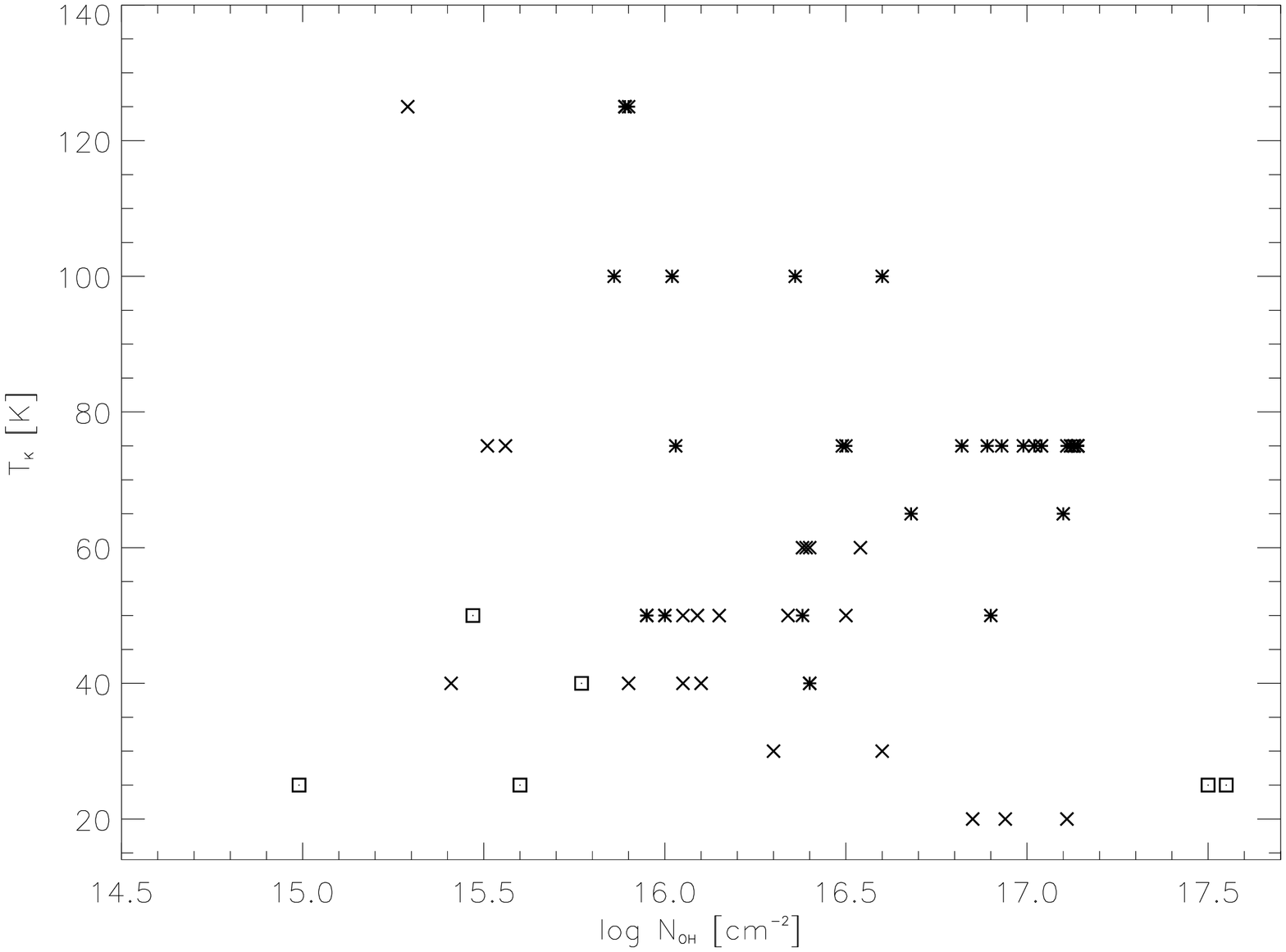}\\
\includegraphics[scale=0.41,angle=90]{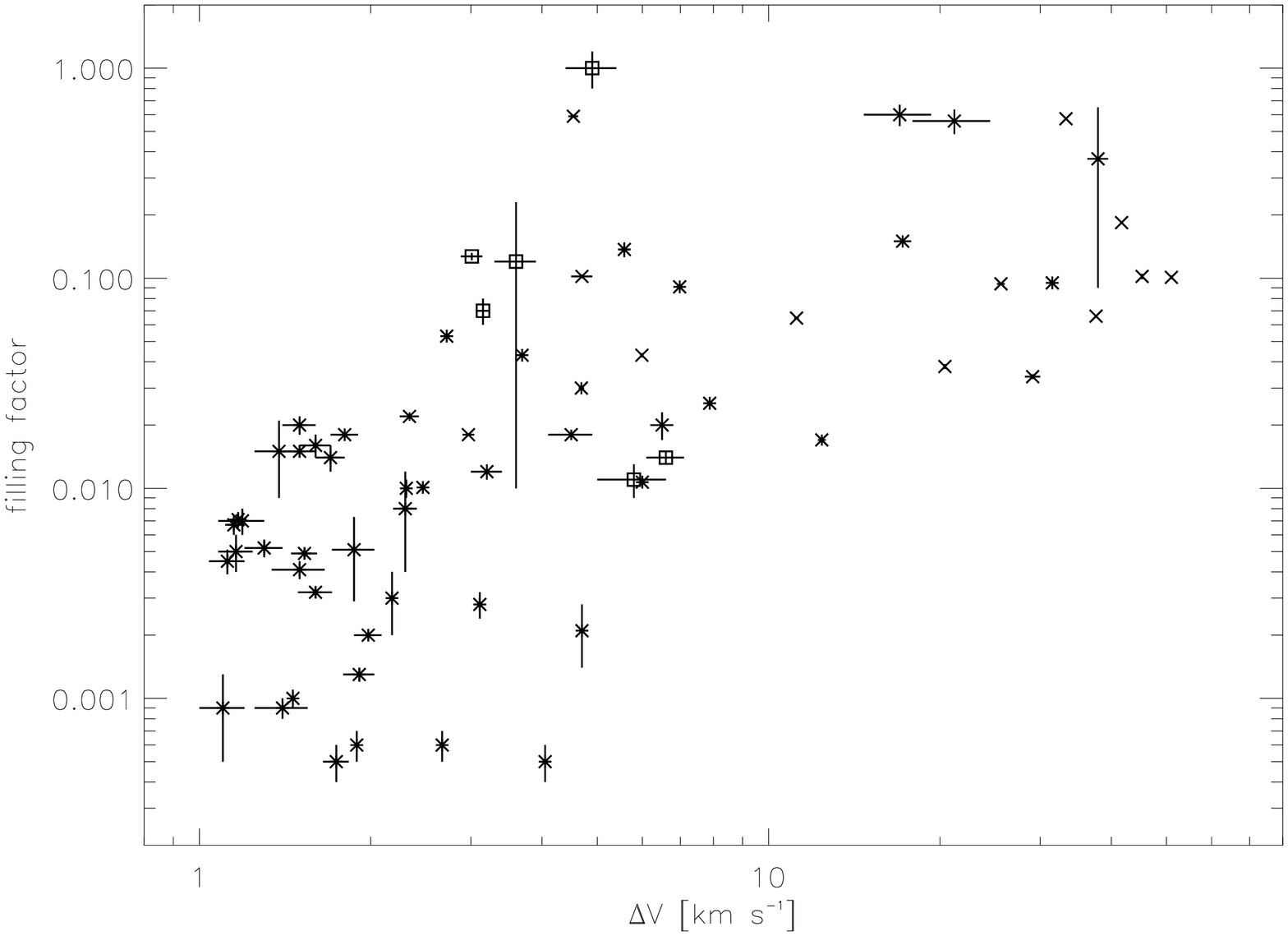}
\caption{Plots of physical quantities from our fitted values: [Top] Hydroxyl column density $\noh$ is plotted against filling factor $f$ with 1$\sigma$ errors. [Center] Hydroxyl column density $\noh$ is plotted against kinetic temperature. [Bottom] Filling factor $f$ is plotted against line width $\Delta$V with 1$\sigma$ errors.   In all plots 
asterisks indicate OH(1720 MHz) maser emission associated with interacting supernova remnants; crosses indicate main-line OH absorption features; and open squares indicate anomalous clouds.
The sensitivity of our observations is biased against detecting lines of both low filling factor and low OH column.\label{fig:logplots}} 
\end{figure}

\end{document}